\documentclass[english]{elsarticle}
\usepackage[T1]{fontenc}
\usepackage[latin9]{inputenc}
\usepackage{array}
\usepackage{textcomp}
\usepackage{amsmath}
\usepackage{graphicx}
\usepackage{setspace}
\usepackage{subscript}
\makeatletter

\providecommand{\tabularnewline}{\\}
\newcommand{\lyxdot}{.}

\makeatother

\usepackage{babel}
\begin{document}

\title{Ability of the e-TellTale sensor to detect flow features over wind
turbine blades: flow separation/reattachment dynamics}

\author{A. Soulier\textsuperscript{1,2}, C. Braud\textsuperscript{2}, D.
Voisin\textsuperscript{1}, B. Podvin\textsuperscript{3}}

\address{1- Mer Agitée, Route de Port-la-Forêt, 29940 La Forêt-Fouesnant}

\address{2- LHEEA (CNRS/ECN), Ecole Centrale Nantes 1, rue de la Noë, 44321
Nantes}

\address{3- LIMSI (CNRS), Campus Univ. bât. 507, Rue John Von Neumann, 91400
Orsay}
\begin{abstract}
Monitoring the flow features over wind turbine blades is a challenging
task that has become more and more crucial. This paper is devoted
to demonstrate the ability of the e-TellTale sensor to detect the
flow separation/reattachment dynamics over wind turbine blades. This
sensor is made of a strip with a strain gauge sensor at its base.
The velocity field was acquired using TR-PIV measurements over an
oscillating thick blade section equipped with an e-TellTale sensor.
PIV images were post-processed to detect movements of the strip, which
was compared to movements of flow. Results show good agreement between
the measured velocity field and movements of the strip regarding the
separation/reattachment dynamics. 
\end{abstract}

\maketitle


\section{Introduction }

Wind turbines are placed in the low layers of the atmospheric boundary
layer where the wind is strongly influenced by the surface roughness
and the thermal stability which creates turbulence and vertical gradients
of the wind \citep{emeis_wind_2018}. The rotor yaw and the blade
pitch alignment within this highly unsteady wind inflow is a subject
that is becoming more and more crucial with the rotor blade lengths
that are increasingly long (107m for the largest existing turbine:
Heliade-X). Also, offshore turbines are arranged in an array layout
and not just in-line, which induces additional sheared inflow conditions
and additional small turbulent structures \citep{chamorro_evolution_2012}.
This results in strong and local variations of speed and directions
on the wind turbine rotor blades. These variations lead to unsteady
aerodynamic effects with turbulent inflows responsible of more than
65\% of fatigue loads \citep{rezaeiha_fluctuations_2017}. To alleviate
these loads, smartblades and/or fluidic actuators are nowadays considered
\citep{pechlivanoglou_passive_2013,jaunet_experiments_2018,batlle_airfoil_2017}.
For this last strategy or to perform blade remote monitoring, one
key issue is the development of robust technologies that are able
to provide an instantaneous detection of the state of the flow on
the blade aerodynamic surface. On current operating wind turbines
the wind is generally monitored using an anemometer situated on the
nacelle. It provides a slow measure of the wind which is perturbed
by the rotor and the nacelle. Moreover being only a one-point measurement,
it does not appreciate shear, yaw/pitch misalignments or turbulence
on blades. Recent monitoring technologies allow to overcome some of
these drawbacks. Among the most mature technologies, the spinner anemometer
is measuring the wind in front of the rotor, removing perturbations
from the rotor \citep{pedersen_aerodynamics_2007}. Also, capabilities,
costs and integration of nacelle-mounted LIDAR, measuring the wind
inflow few diameters upstream of the rotor, have been significantly
improved during the last decades \citep{aubrun_wind_2016} \citep{bossanyi_wind_2014}.
However, from the knowledge of the authors, nothing is yet able to
measure the state of the flow on current blades. Some field measurement
campaigns were punctually performed for research purposes using pressure
probes around the blades on dedicated blade manufacturing with however
only weak potential of these sensors to be used in a day-to-day operation
of wind turbines \citep{troldborg_danaero_2013}. Some solutions were
explored such as tufts or stall flags glued on the blade correlated
with positions of the flow separation \citep{swytink-binnema_digital_2016,pedersen_using_2017}\citep{corten_flow_2001}.
However, these methods need a mounted camera on the turbine with its
associated drawbacks (fragility of the camera, vision at night ...). 

\begin{figure*}
\begin{centering}
\begin{tabular}{cc}
a)\includegraphics[width=0.5\textwidth]{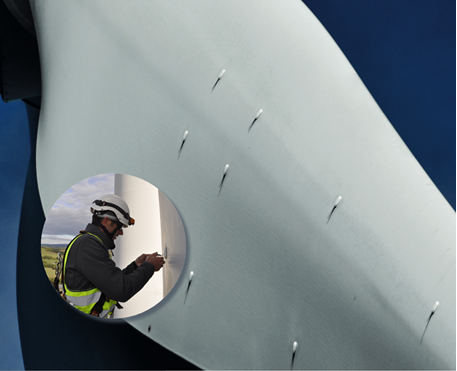} & b)\includegraphics[width=0.5\textwidth]{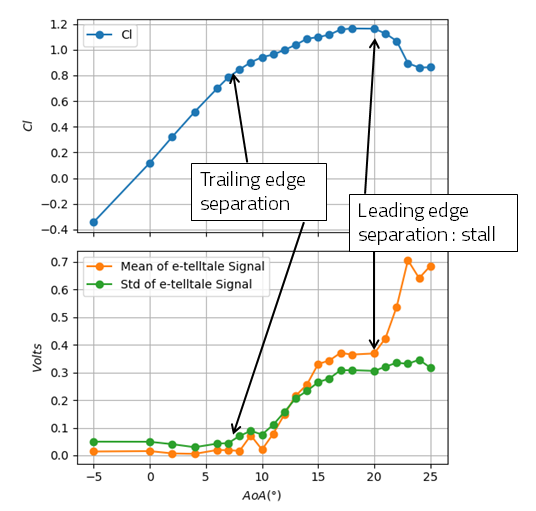}\tabularnewline
\end{tabular} 
\par\end{centering}
\caption{Previous studies: a) Robustness and practical mounting issues solve
on EDF-Renewable wind turbines b) Ability of full scale e-TellTale
sensors to detect static flow separations at high chord Reynolds numbers
($10^{6}$) from wind tunnel tests \label{fig:epenon-avant}}
\end{figure*}

An interesting alternative to these technologies is the electronic
telltale sensor, developed by Mer Agitée\footnote{https://www.meragitee.com/}.
It is composed of a strip moving like a tuft but with a strain gauge
encased in its base making it able to transmit the information directly
to any monitoring or control system through an embedded wireless electronic
unit. It has been originally developed to detect flow separation on
sails of offshore racing sailing vessels and has been recently adapted
for wind turbine blade monitoring. Robustness and practical mounting
issues were solved from industrial tests (figure \ref{fig:epenon-avant}a),
while full scale tests of the device were performed at high Reynolds
numbers in the wind tunnel facility of CSTB to demonstrate the relation
between the e-Telltale sensor signal and the lift curve for static
variations of the angle of incidence as can be seen in figure \ref{fig:epenon-avant}b
\citep{soulier_electronic_2017}. The present study is intended to
study the ability of the e-TellTale sensor to dynamically detect the
flow separation/reattachment phenomena. For that purpose, experiments
at a downscaled 2D blade section were performed in the LHEEA aerodynamic
wind tunnel, using Time Resolved PIV and different post-processing
methods to extract the strip position of the sensor in the flow field
(vision algorithms) and to evaluate the flow separation over the aerodynamic
surface (POD, and vortex detection). The experimental set-up and the
post-processing methods are described in paragraph 2 and 3 respectively.
Results are presented in the 4th paragraph including: a description
of the baseline flow (4.1), results of the different post-processing
methods to detect the flow separation (4.2), results on the ability
of the e-TellTale sensors to detect flow separation (4.3). 

\section{Experimental Setup}

\label{sec:wind_tunnel_facility}

The experiments were performed in the recirculating aerodynamic wind
tunnel facility of the LHEEA laboratory at Centrale Nantes (France).
The working section is 0.5x0.5m\texttwosuperior{} and 2.4m long with
a turbulent intensity less than 0.3\% of turbulence. The Reynolds
number based on the chord length of the 2D blade section, c$\simeq$0.09m,
is $Re_{c}=(U_{\infty}c)/\nu\simeq2.10^{5}$ with $U_{\infty}=35$m/s
the free-stream velocity. 

\subsection{Blade profile}

\label{subsec:blace_profile}

\begin{figure*}
\begin{centering}
\includegraphics[width=0.5\textwidth]{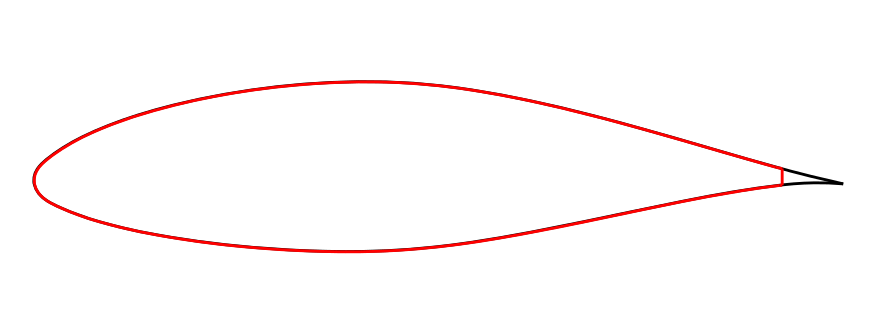}
\par\end{centering}
\caption{ NACA 65\protect\textsubscript{4}-421profile manufactured in red
and the theoretical trailing edge in black \label{fig:modif_NACA}}
\end{figure*}

Measurements were performed using a NACA 65-421 profile in composite
material. Due to the fabrication process, it is truncated at 91\%
of the chord length so that the trailing edge thickness is 2 mm (see
figure \ref{fig:modif_NACA}). A similar profile was already used
by Jaunet \& Braud\citep{jaunet_experiments_2018} to demonstrate
the ability of local micro-jets to alleviate loads. It is a thick
profile with two drops on the lift coefficient curve corresponding
to a first boundary layer separation at the trailing edge of the profile
for AoA\textasciitilde{}8\textdegree , and a second flow separation
at the leading edge for AoA\textasciitilde{}20\textdegree{} causing
stall. From 8\textdegree{} to 20\textdegree{} the separation point
moves gradually from the trailing edge to the leading edge, corresponding
to a gradual variation of the loads. 

An oscillating motion was imposed using a crank drive for the linear
movement imposed by a feedback linear motor from LinMot. This oscillating
motion was checked from PIV image processing using the detection of
the blade surface at the position of the e-Telltale sensor. The detection
of the blade surface was also later used to extract the position of
the e-TellTale sensor in the vector field (see section \ref{subsec:strip detection}).
The amplitude of the blade oscillation, $\varDelta\alpha_{0}$=5\textdegree ,
was chosen so that the flow, initially separated at the trailing edge,
moves gradually towards the leading edge flow separation where the
stall occurs (see PIV vector fields in figures \ref{fig:Instantaneous-velocity-fields}
and \ref{fig:Instantaneous-velocity-fields-1}). The oscillating frequency,
$f_{osc}$=1Hz, was chosen similar to the study of Jaunet \& Braud
\citep{jaunet_experiments_2018} to mimic a constant shear inflow.
This leads to a reduce frequency of $k=\pi f_{osc}c/U_{\infty}$=0.008
corresponding to a quasi-steady stall behavior \citep{choudhry_insight_2014}.
The blade was equipped with a e-telltale sensor at mid-span on the
suction side. Figure \ref{fig:schema_veine}b) shows the e-telltale
on the surface of the 2D blade profile installed in the LHEEA aerodynamic
wind tunnel. A small part ($\simeq5mm$) of the pink strip of the
e-telltale sensor is glued on a strain gauge sensor, itself glued
on a thin stainless steel sheet embedded into the blade. The rest
of the strip is free to move above the aerodynamic surface. Its length
is one third of the blade chord. The signal from the strain gauge
sensor was not acquired during PIV measurements, however, it was checked
that the signal from this strip, made of a nylon fabric, behaves similarly
as full-scale experiments from \citep{soulier_electronic_2017}. In
particular it was checked that we are able to distinguish the level
of the signal when the sensor is in the two different flow states
over the aerodynamic surface: attached / separated. 

\subsection{PIV measurements}

\begin{figure*}
\centering{}%
\begin{tabular}{cc}
a) \includegraphics[width=0.55\textwidth]{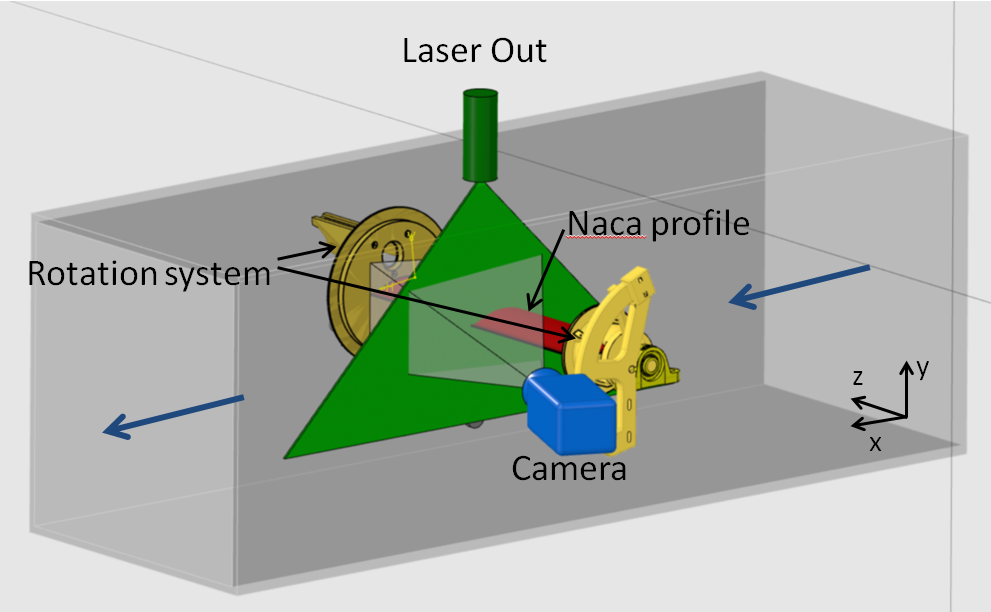} &  b) \includegraphics[width=0.45\textwidth]{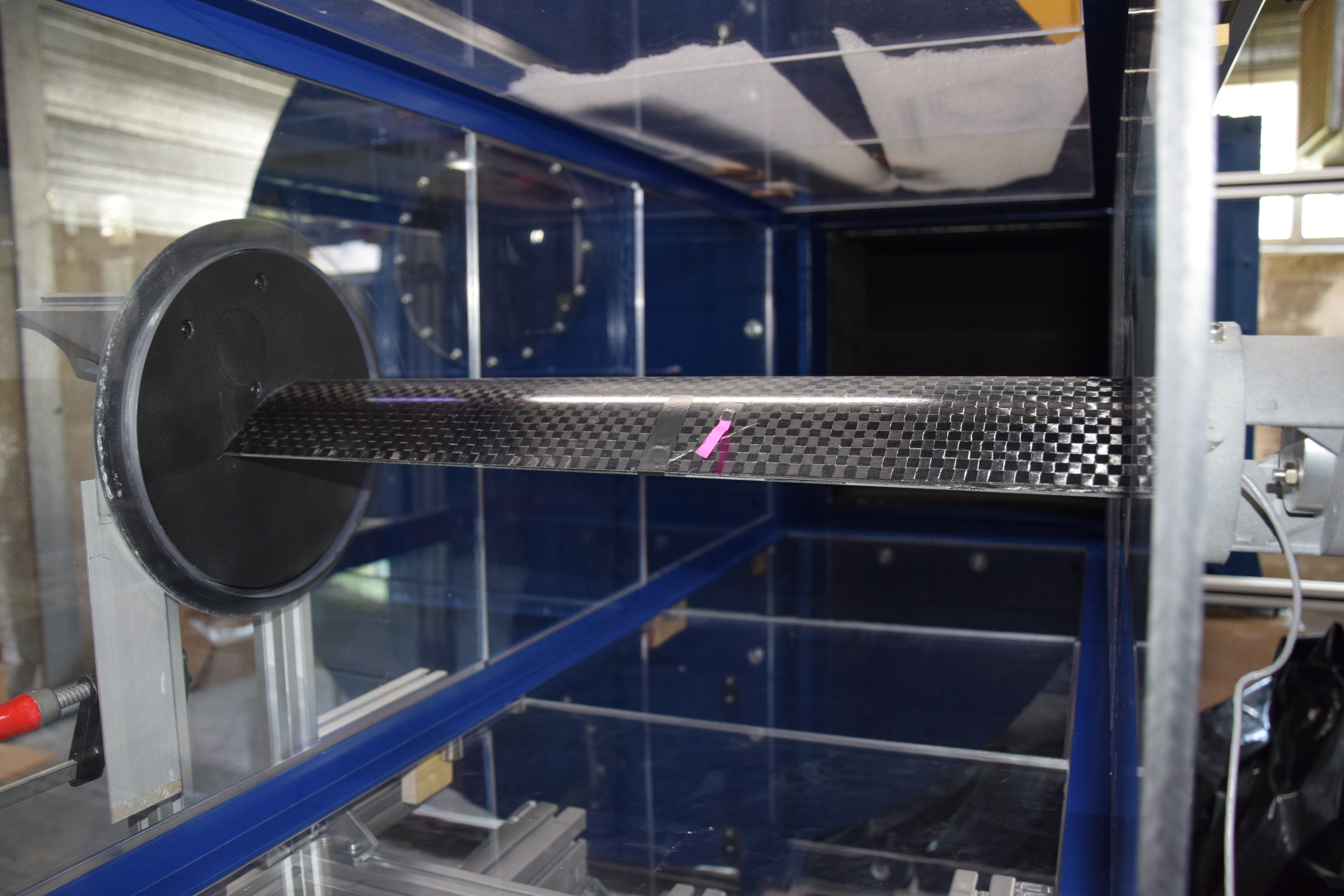}\tabularnewline
\end{tabular}\caption{a) Scheme of the PIV set-up with the axes (x,y,z). b) : The 2D blade
section mounted in the LHEEA aerodynamic wind tunnel with the e-telltale
in pink\label{fig:schema_veine}}
\end{figure*}

Flow data were collected with a TR-PIV system able to produce 1600
velocity fields each second. A DM20-527 DH laser from Photonics Industries
delivering a 2x20 mJ double laser sheet at the green wavelength of
527 nm was used in this setup. The camera was a Phantom Miro M310,
recording 1200 x 800 px\texttwosuperior{} images at 3200 Hz, the 6Gb
of Ram memory of the camera allowed to capture 2000 velocity fields
for each run. The camera was equipped with a Zeiss Makro Planar 2/50
lens (i.e.$f=50mm,\:a=f/2$). With this setup, the field of view was
216 x 106 mm\texttwosuperior{} leading to a spatial resolution of
6.3 px/mm. The PIV velocity fields were computed using a 16 x 16 px\texttwosuperior{}
interrogation area with an overlap of 50\% leading to a grid resolution
of 159 x 99 with a maximum spacing between vectors of 1.3mm or 0.014c.
As seen in the figure \ref{fig:schema_veine} the optical axes of
the camera was not totally perpendicular to the laser sheet. After
calibrating this misalignment by taking snapshots of the calibration
target, all the raw images and the velocity fields were dewarped.
In addition to the classical noise inherent to PIV measurement, the
presence of the e-telltale strip in the field of view of the PIV camera
caused some spurious vectors explained by some light shoots on images
when the clear fabric of the strip reflect the laser light directly
towards the camera. To remove and replace these spurious vectors the
automated post-processing algorithm developed by Garcia \citep{garcia_fast_2011}
was used. 

\section{Introduction in processing Methods}

\subsection{Strip detection method}

\label{subsec:strip detection}

\begin{figure*}
\begin{centering}
\begin{tabular}{cc}
a)~\includegraphics[width=0.5\textwidth]{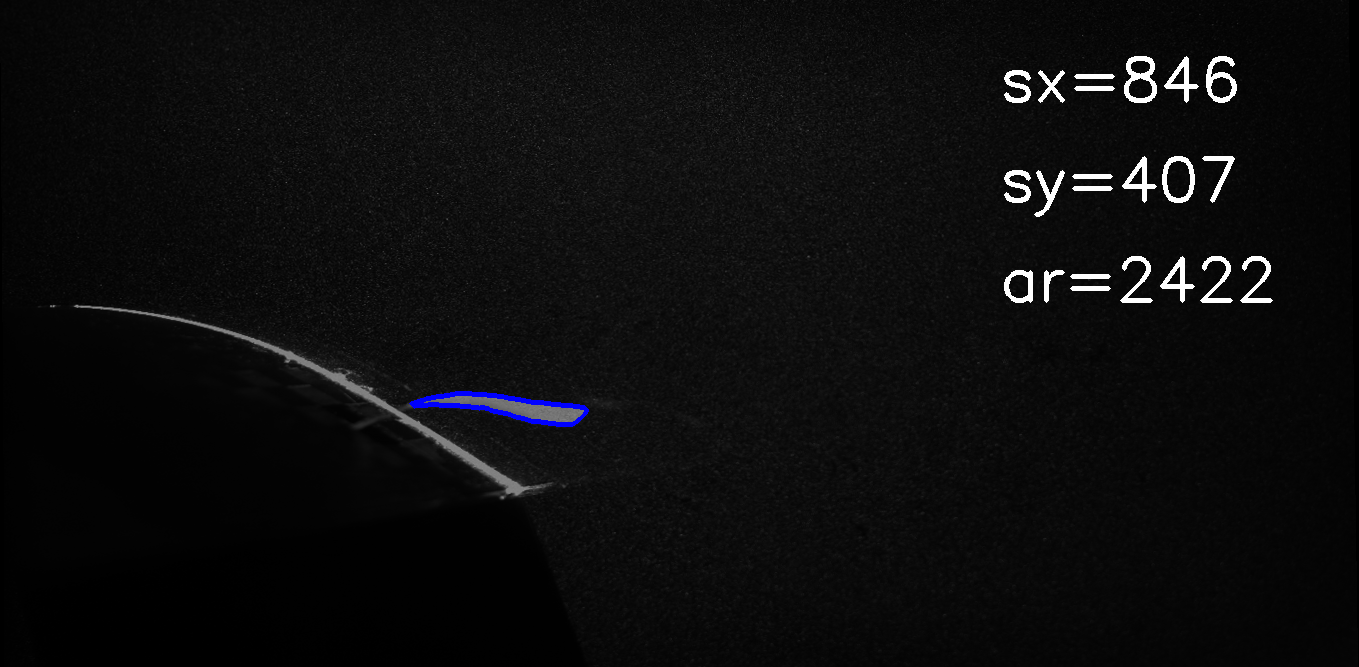} & b) \includegraphics[width=0.5\textwidth]{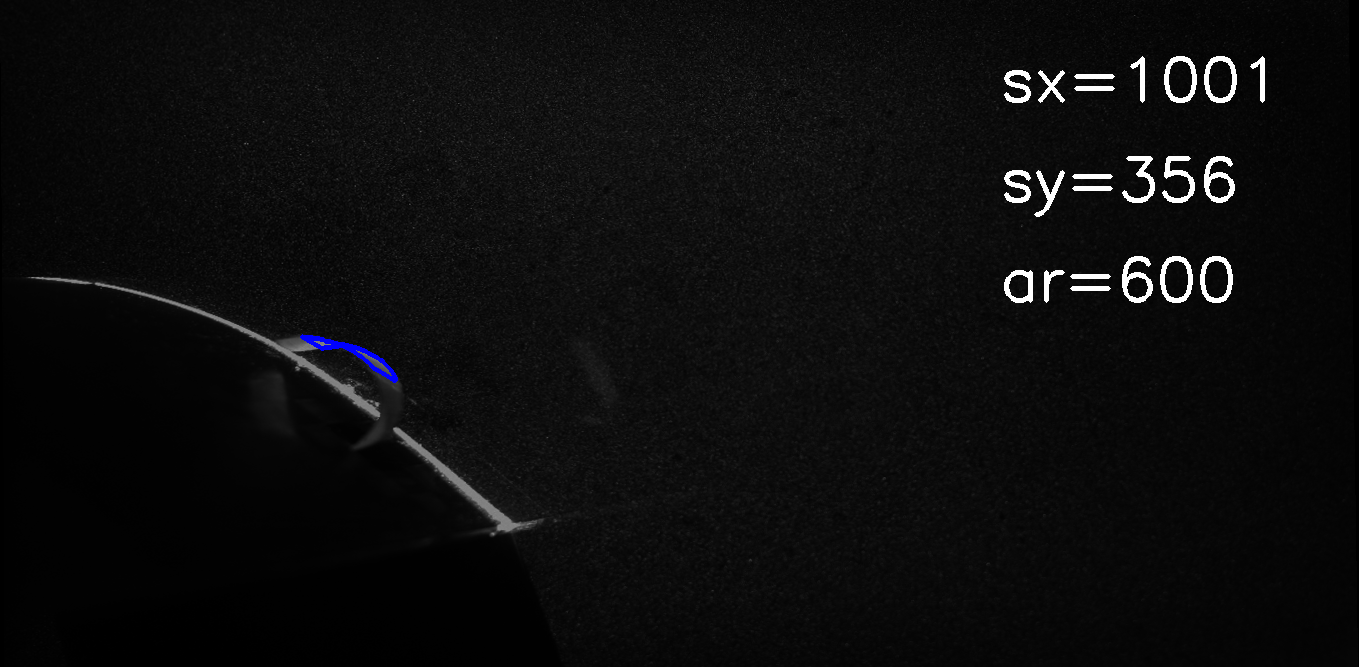}\tabularnewline
\multicolumn{2}{c}{c)\includegraphics[width=0.5\textwidth]{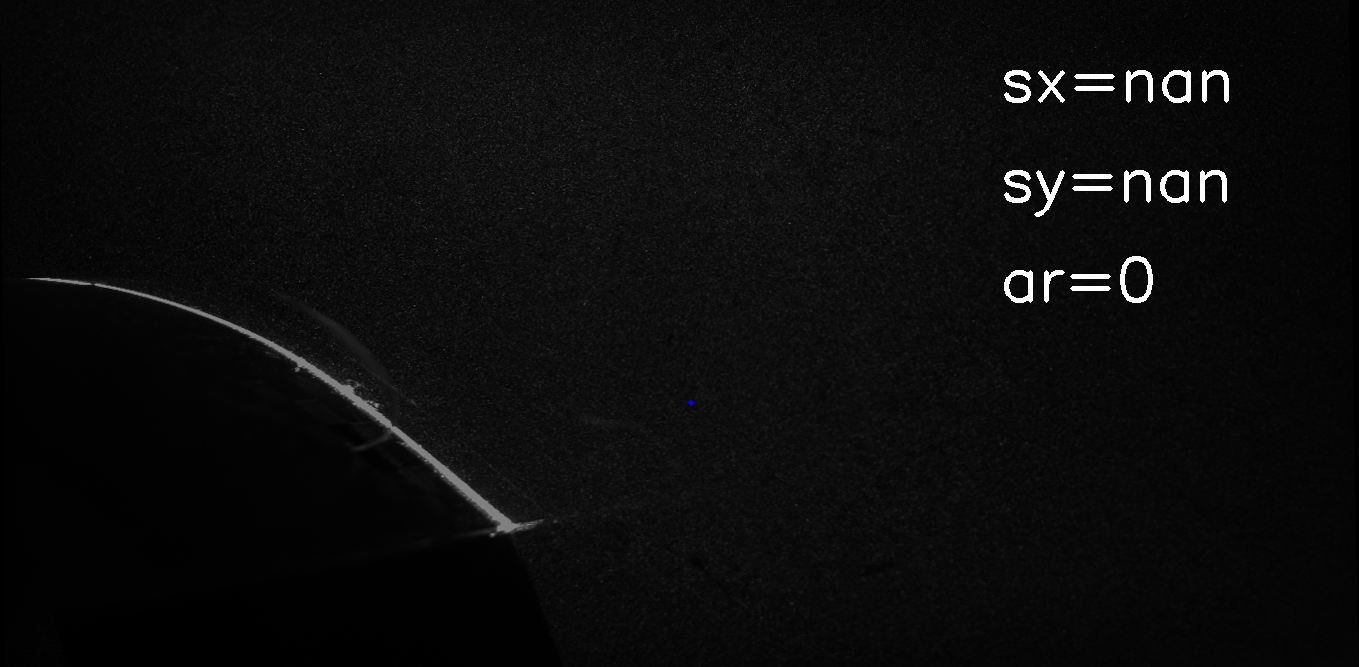}}\tabularnewline
\end{tabular} 
\par\end{centering}
\caption{Detected strip contour from PIV images using OpenCV: a) for the attached
flow case and b) for the detached flow case c) corresponding to an
outlier case (impossible to detect the strip position). sx and sy
are respectively ~the streamwise and spanwise positions in pixels
of the center of the detected area (in blue). ar is the area of the
detected contour in pixels \label{fig:strip_ID}}
\end{figure*}

\begin{figure}
\begin{centering}
\begin{tabular}{cc}
a) & \includegraphics[width=0.7\textwidth]{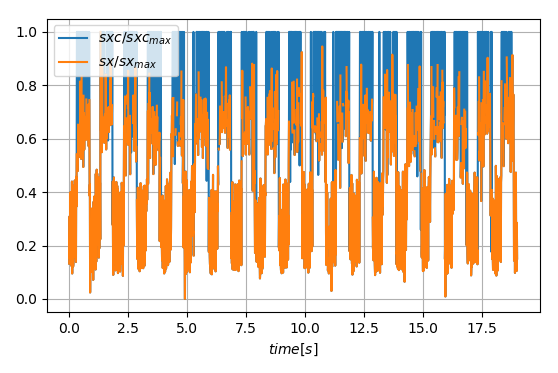} \tabularnewline
b) & \includegraphics[width=0.7\textwidth]{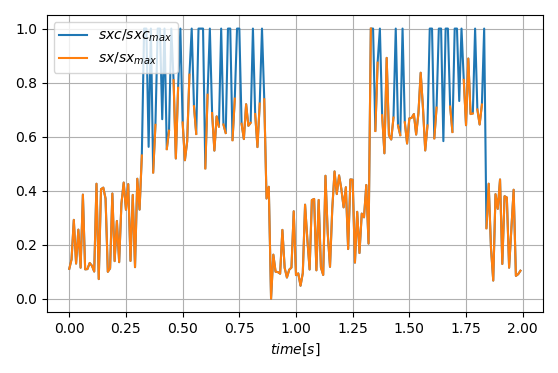} \tabularnewline
\end{tabular}
\par\end{centering}
\caption{Streamwise coordinate of the identified strip $sx$ before correction
and $sxc$ after correction a) the full run and b) a zoom on the two
first oscillations \label{fig:sx_brut}}
\end{figure}

The flow field over the aerodynamic surface is measured using TR-PIV
measurements during the oscillations of the blade profile. To extract
movements of the e-TellTale strip within this flow field, PIV images
were post-processed using vision algorithms from the Open Source Computer
Vision Library\footnote{http://opencv.org} . The chosen methodology
uses PIV images containing laser reflections of the blade surface
and of the strip. The first step is to separate the blade surface
contour from the strip contour. The images were first binarized so
that white pixels, corresponding to the reflection of the laser on
the blade and the strip surfaces, are set to 1 and all others to 0.
To separate pixel coordinates of the blade from pixel coordinates
of the strip, a local gradient of white pixel coordinates is computed,
revealing ordinates of pixels corresponding to the strip location.
Then, the resulting curve was smoothed using a Savitzky-Golay filter.
Finally, this resulting identified profile curve was fit to the theoretical
suction side profile curve to extract the best euclidean transformation
(i.e. only rotation, translation and uniform scaling considered for
the transformation) going from the measured curve to the theoretical
profile. This was done using a function of OpenCV which primarily
uses the RANSAC algorithm to detect spurious points and then the Levenberg-Marquardt
algorithm to fit the profile. The result is a transformation matrix
from which an angle of rotation is extracted. Also, from the detected
blade surface contour, a mask is defined to remove everything below
it so that the remaining bright contour is the strip. The resulting
cleaned binarized images were then used to extract the strip location
using a contour detection function from OpenCV. The contour detection
function recognize the white pixels surrounded by other white pixels
and regroup all of it in one entity. As we are interested in the flow
separation phenomena over the aerodynamic surface which induces large
movements of the strip from the downstream to the upstream flow direction,
it was found sufficient to resume the position of the strip by the
center position of the detected contour. The strip detection method
was first checked visually on some samples such as the figure \ref{fig:strip_ID}
which shows raw PIV images on which the detected aera is circled in
blue with the coordinate of its center noted $sx$ and $sy$ for the
respective streamwise and spanwise directions. It was then possible
to automatize the method for images of the oscillating blade periods.
Missing values present in the signal are related to default in the
contour detection algorithm as can be seen in the figure \ref{fig:strip_ID}c.
These outliers are found to be correlated with AoA beyond stall, were
3D effects are dominants. These values were replaced by the maximum
value of $sx$. The corrected signal, $sxc$, is presented with the
original signal $sx$ in the figure \ref{fig:sx_brut}. 

\subsection{Vortex identification method}

\label{subsec:vortexID}

\begin{figure}
\begin{centering}
\includegraphics[width=0.7\textwidth]{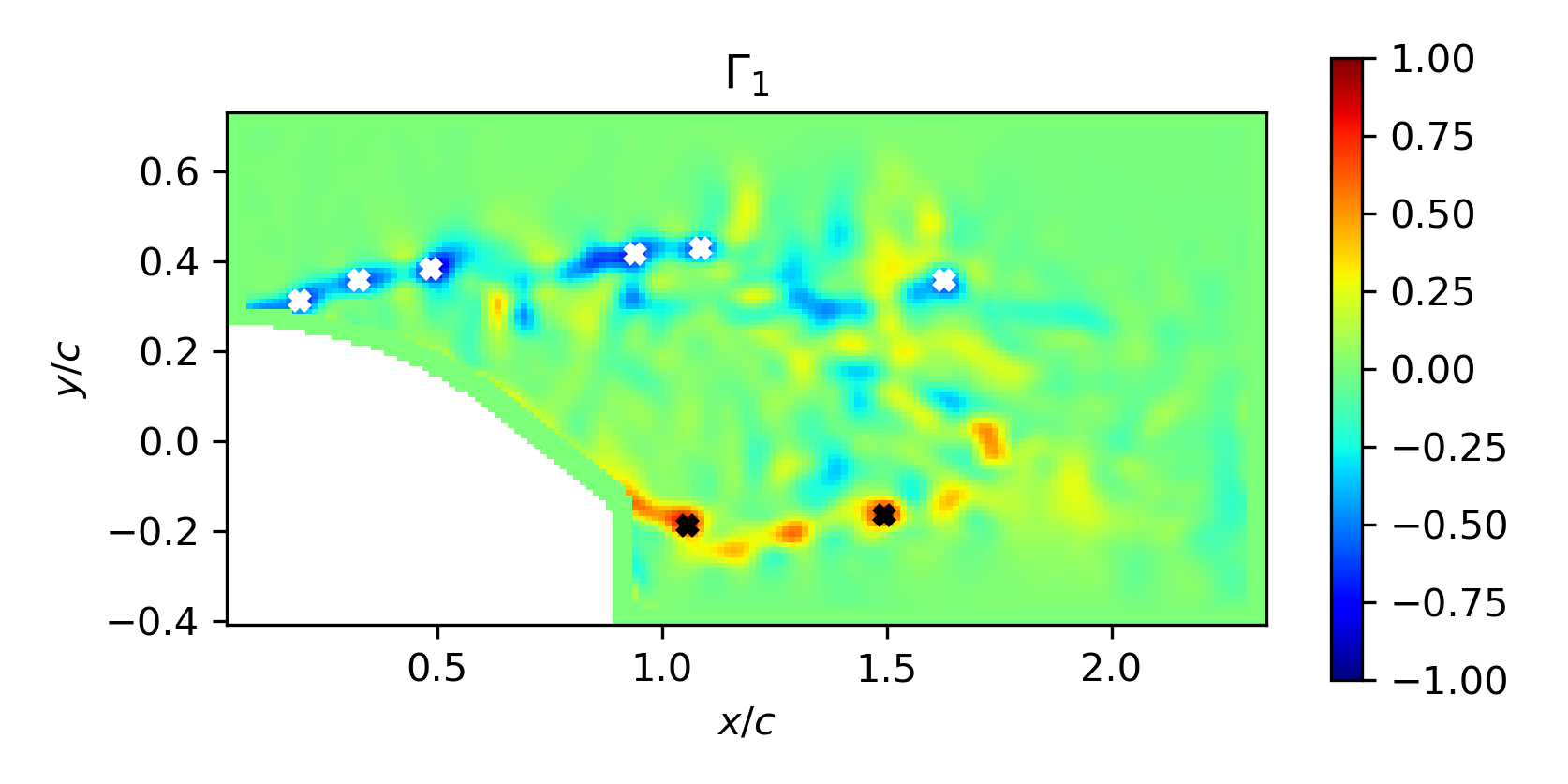} 
\par\end{centering}
\caption{Instantaneous isocontour map of the $\Gamma_{1}$ field with peaks
identified using white cross markers for clockwise vortices and black
cross markers for anticlockwise vortices \label{fig:gamma_1-1}}
\end{figure}

Vortex identification methods are widely spread in the literature
(see e.g. \citep{jeong_identification_1995}). As they enable to distinguish
swirling motion from shearing motion, they were developed to help
in the understanding of turbulent flows and more recently as a real-time
processing method for flow control purposes (see e.g. \citep{braud_real-time_2018}).
In the present study, the $\Gamma_{1}$ criterion method is used \citep{michard_identification_1997}.
This is a geometrical criterion defined as follows: 

\begin{equation}
\Gamma_{1}(P)=\frac{1}{N}\sum_{S}\frac{(PM\land U_{M}).z}{\Vert PM\Vert.\Vert U_{M}\Vert}
\end{equation}
 where $N$ is the number of $M$ points of the square area $S$ around
the center $P$, $U_{M}$ the velocity at the point $M\in S$ and
$z$ the normal unit vector. The size of$S$ act as a spatial filter
. For this study different sizes of $S$ from 9 to 3 grid points were
tested and the differences were found not significant. The presented
results were obtained with $S$ being a square of 7 points. From this
definition, $\Gamma{}_{1}$ is a dimensionless scalar ranging from
$-1$ to $1$, which local extremum indicates the center of a vortex.
Compared to other methods such as the well known Q criteria, the $\Gamma_{1}$
criteria provides equivalent results, with the advantages to avoid
computation of gradients (i.e. decreasing noise) and to provide the
sign of vortices. Similarly as Mulleners and Raffel (2013) \citep{mulleners_dynamic_2013},
the vortex identification method was used to extract vortex locations
in the shear areas over the blade surface during the blade oscillation
cycles (see figure \ref{fig:gamma_1-1}for an illustration of an instantaneous
$\Gamma_{1}$ field).

\subsection{Proper Orthogonal Decomposition }

\label{subsec:POD}

The Proper Orthogonal Decomposition (POD), is a statistical technique
\citep{holmes_turbulence_1996} that extracts spatial modes $\underline{\Psi}(\underline{x})$
that are best correlated on average with a given field $\underline{u}(\underline{x},t)=(u,v)$
defined on a domain $\Omega$. Let $<.>$ denote the temporal average.
The field $\underline{u}(\underline{x},t)$ can be written as a superposition
of spatial modes whose amplitude varies in time 
\[
\underline{u}(\underline{x},t)=<\underline{u}(\underline{x},t)>+\sum_{n}a^{n}(t)\underline{\Psi}^{n}(\underline{x})
\]
The modes can be identified with the method of snapshots \citep{sirovich_turbulence_1987},
which is based on the computation of the temporal autocorrelation
$C$ for a given set of $N$ snapshots $\underline{u}(\underline{x},t_{i}),i=1,\ldots N$:
\[
C_{nm}=\int_{\Omega}\tilde{\underline{u}}(\underline{x},t_{n})\tilde{\underline{u}}(\underline{x},t_{m})d\underline{x},
\]
 where $\underline{u}$ represents the fluctuating part of the snapshots
($\tilde{\underline{u}}(\underline{x},t_{n})=\underline{u}(\underline{x},t_{n})-$
$<\underline{u}(\underline{x},t)>$). The temporal amplitudes are
eigenfunctions of 
\[
C_{nj}a^{p}(t_{j})=\lambda^{p}a^{p}(t_{n})
\]
They are uncorrelated and their variance is given by
\[
<a^{n}a^{m}>=\lambda^{n}\delta_{nm}.
\]
The spatial modes are then obtained from
\[
\underline{\Psi}^{n}(\underline{x})=\sum_{i=1}^{N}a^{n}(t_{i})\underline{u}(\underline{x},t_{i}).
\]
By construction, the modes are orthonormal
\[
\int_{\Omega}\underline{\Psi}^{n}(\underline{x}).\underline{\Psi}^{m}(\underline{x})d\underline{x}=\delta_{nm}.
\]
POD was applied to the 2-D PIV vector fields over two different domains.
The largest domain is used in the description of the baseline flow
(section \ref{subsec:baseline_flow}), while the smaller domain is
used to detect the flow separation/reattachment dynamics in the oscillating
cycle (see section \ref{subsec:strip detection}) 

\section{Results}

Results are presented in three steps. Firstly, the baseline flow is
described, including a description of the flow during an oscillation
cycle and the description of the secondary oscillation in the wake
flow when separated. Then, three methods to detect the flow separation
from PIV measurements are presented and compared. At last, results
of the detection of the strip are compared to these methods to evaluate
the ability of sensor to detect the flow separation.

\subsection{The baseline flow }

\label{subsec:baseline_flow}
\begin{figure}
\begin{centering}
\includegraphics[width=0.7\textwidth]{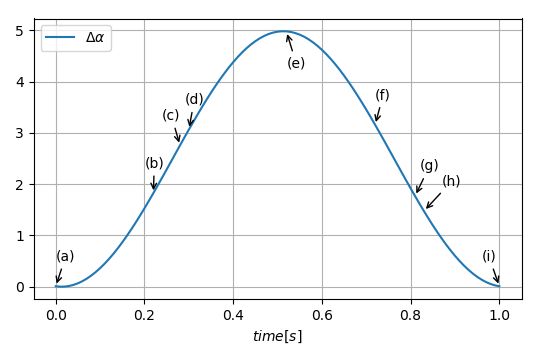} 
\par\end{centering}
\caption{Evolution of the relative angle of attack $\Delta\alpha$. (x) : instantaneous
velocity fields detailed below \label{fig:alpha_evolv}}
\end{figure}

\begin{figure*}
\begin{centering}
\begin{tabular}{cc}
\multicolumn{2}{c}{\includegraphics[width=0.5\textwidth]{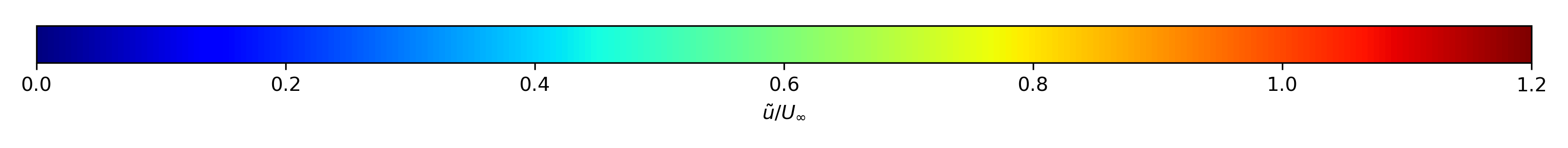}}\tabularnewline
\includegraphics[width=0.5\textwidth]{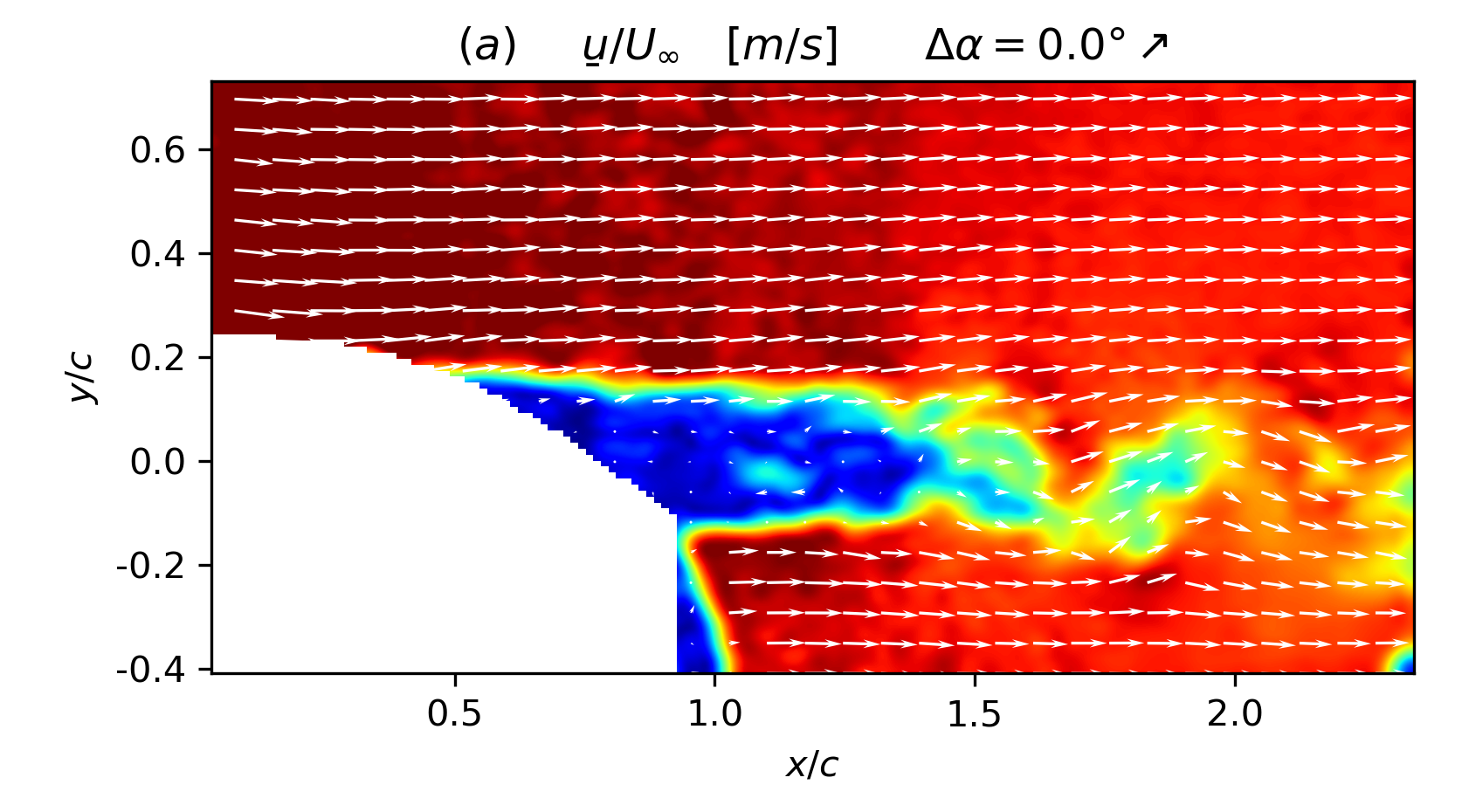} & \includegraphics[width=0.5\textwidth]{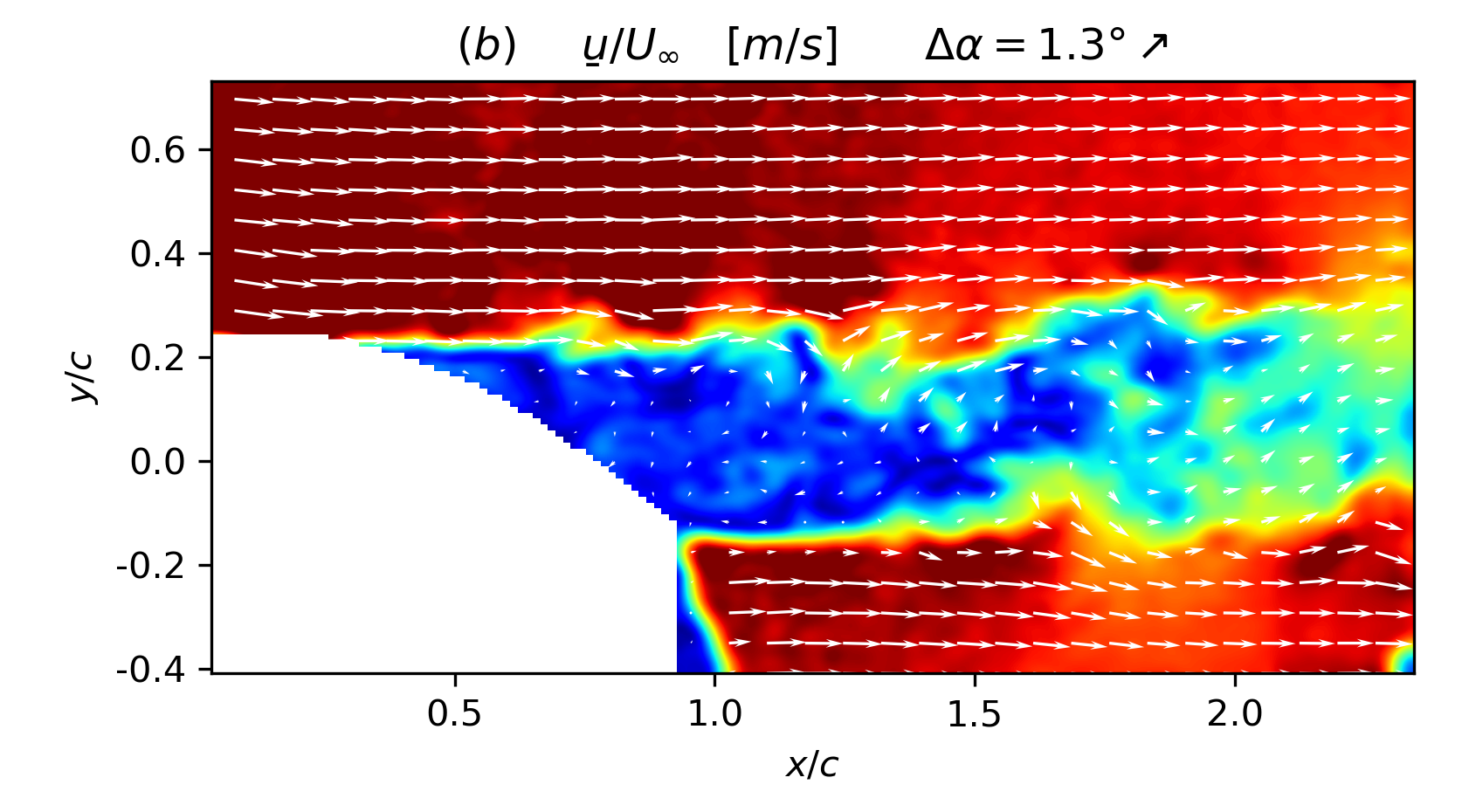}\tabularnewline
\includegraphics[width=0.5\textwidth]{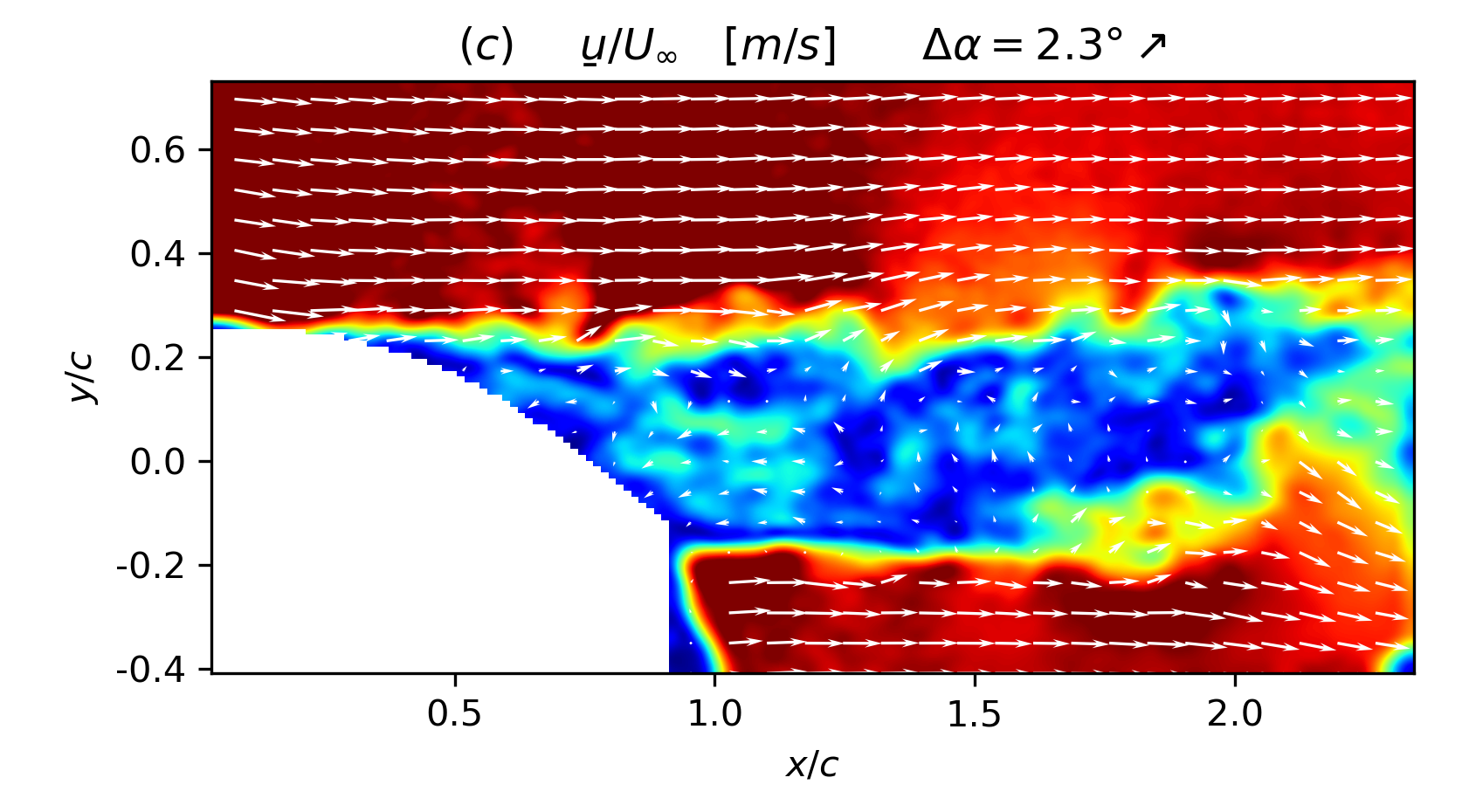} & \includegraphics[width=0.5\textwidth]{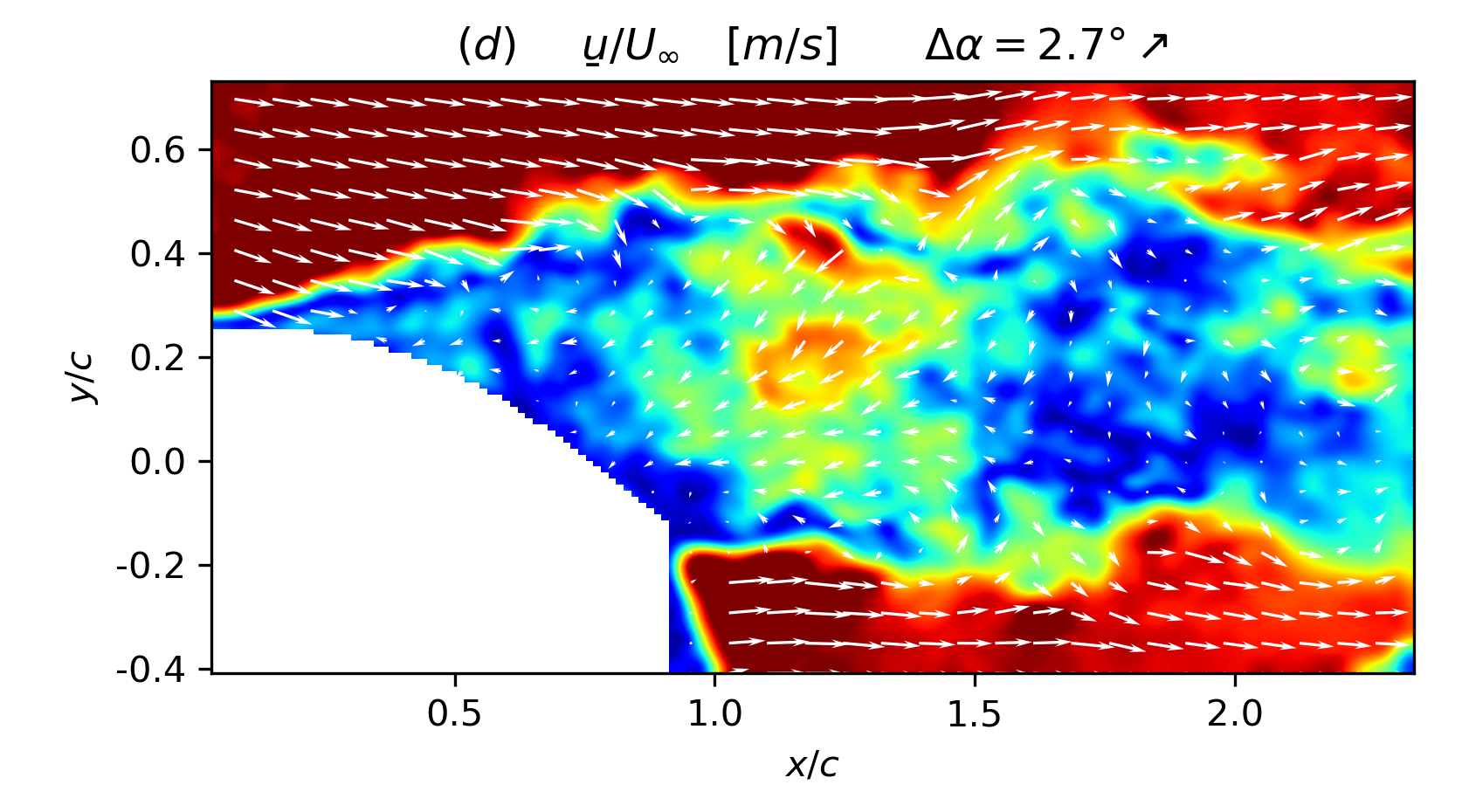}\tabularnewline
\multicolumn{2}{c}{\includegraphics[width=0.5\textwidth]{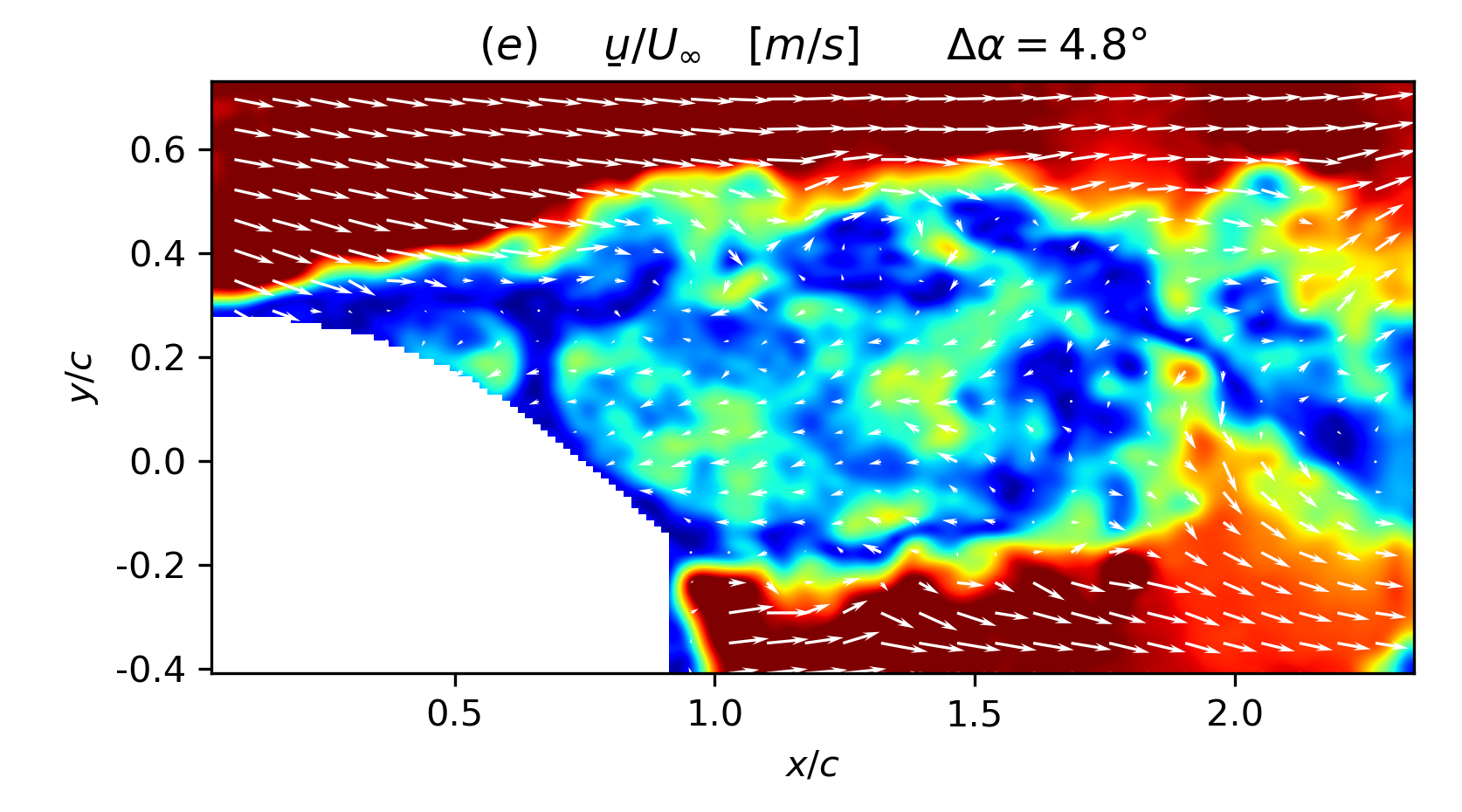}}\tabularnewline
\end{tabular}
\par\end{centering}
\begin{centering}
~
\par\end{centering}
\begin{centering}
\par\end{centering}
\caption{Instantaneous velocity fields superposed with isocontours of the velocity
modulus (i.e. $\sqrt{{\parallel u\parallel^{2}+\parallel v\parallel^{2}}}$
) at different $\Delta\alpha$ corresponding to points of the blade
oscillation given in figure \ref{fig:alpha_evolv} during the upstroke
phase (noted $\nearrow$): (a) is a point at the lowest $\Delta\alpha$
of the upstroke phase of the oscillation cycle, (b) is an intermediate
point, (c) is a point just prior to full stall angle, (d) is a point
just after the stall angle and (e) corresponds to a point at the maximum
amplitude of the blade oscillation cycle\label{fig:Instantaneous-velocity-fields}}
\end{figure*}

\begin{figure*}
\begin{centering}
\includegraphics[width=0.5\textwidth]{V35_AoA718_3Hz_Penon2_Run_Num_11CB}
\par\end{centering}
\begin{centering}
\begin{tabular}{cc}
\includegraphics[width=0.5\textwidth]{UV_t_754} & \includegraphics[width=0.5\textwidth]{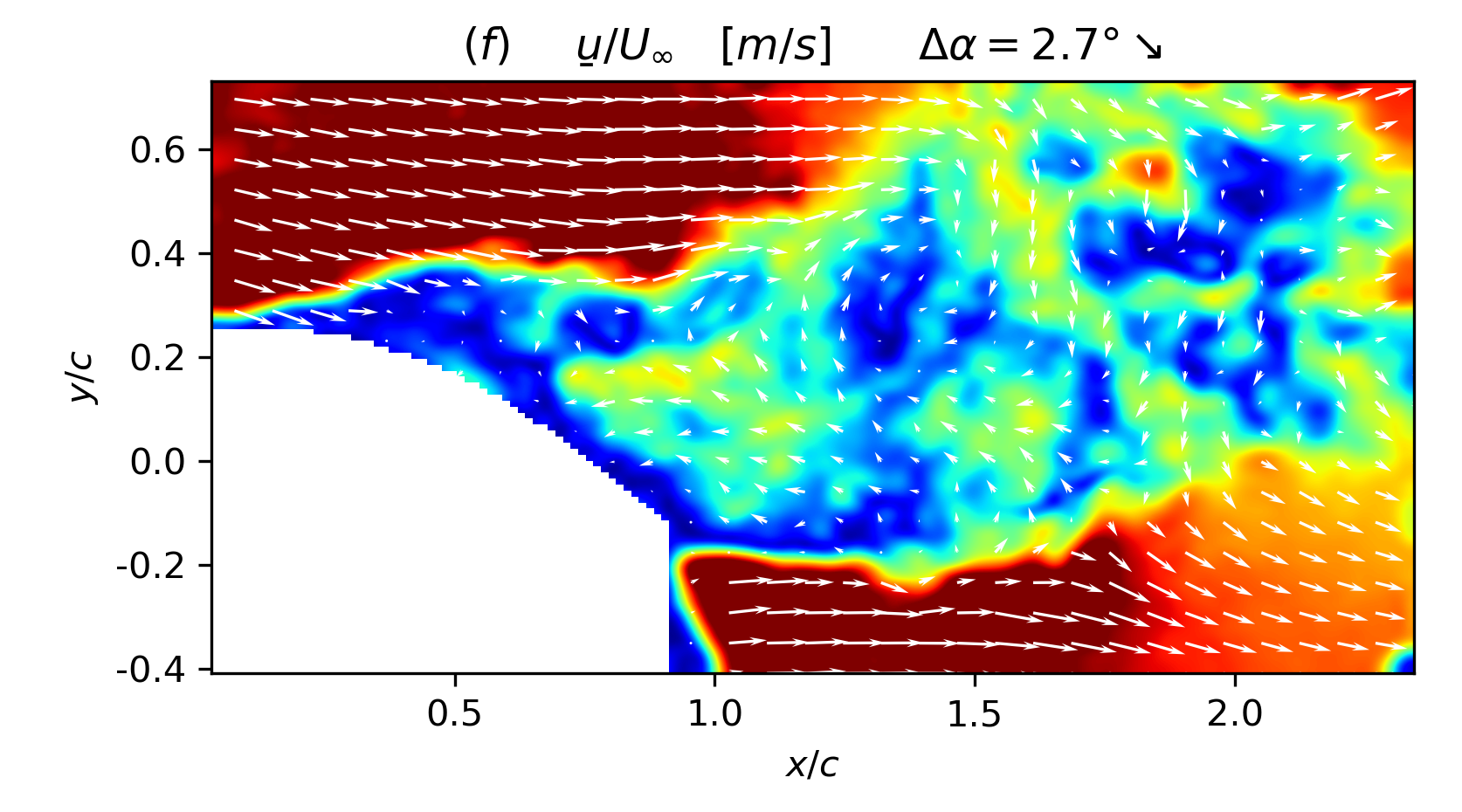}\tabularnewline
\includegraphics[width=0.5\textwidth]{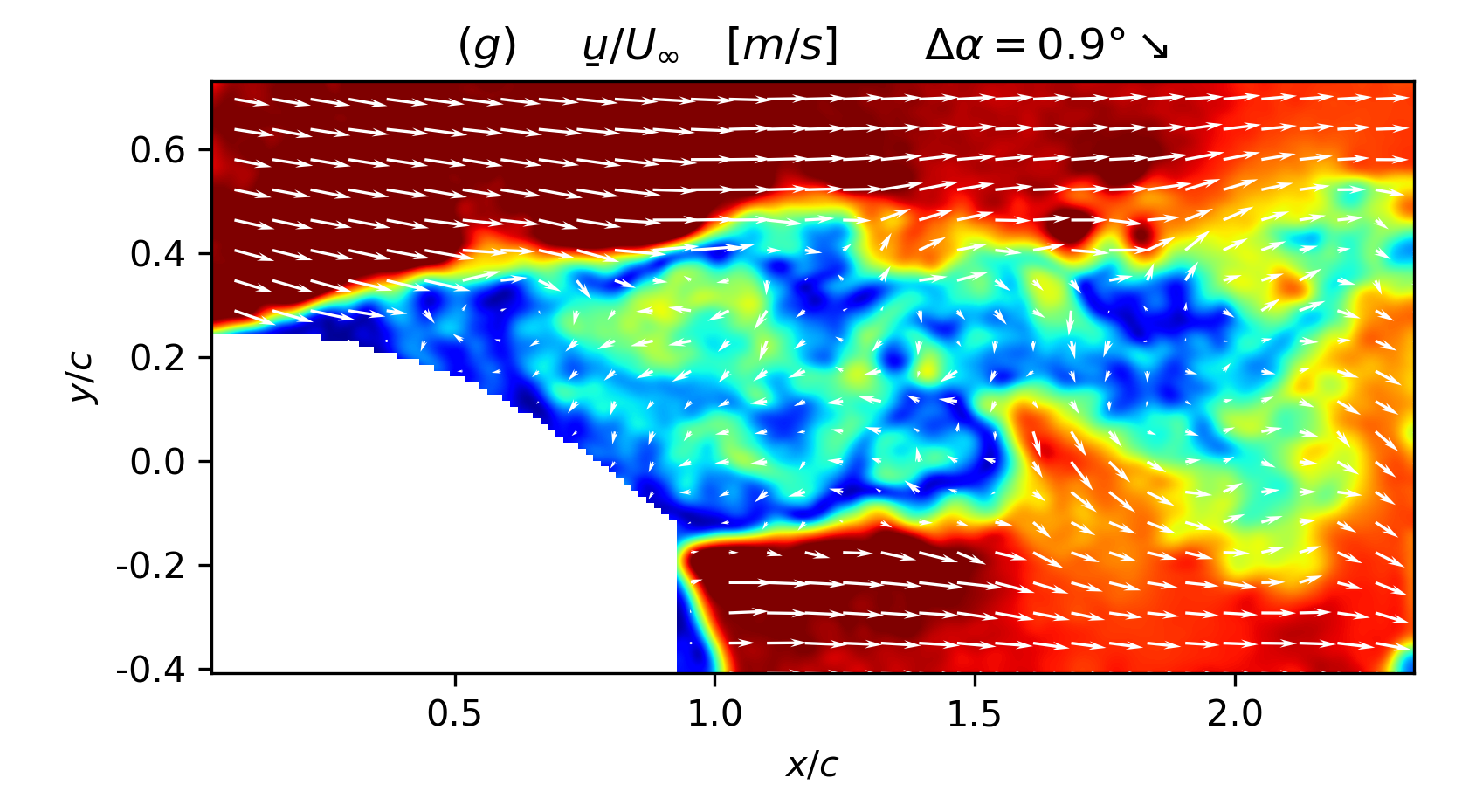} & \includegraphics[width=0.5\textwidth]{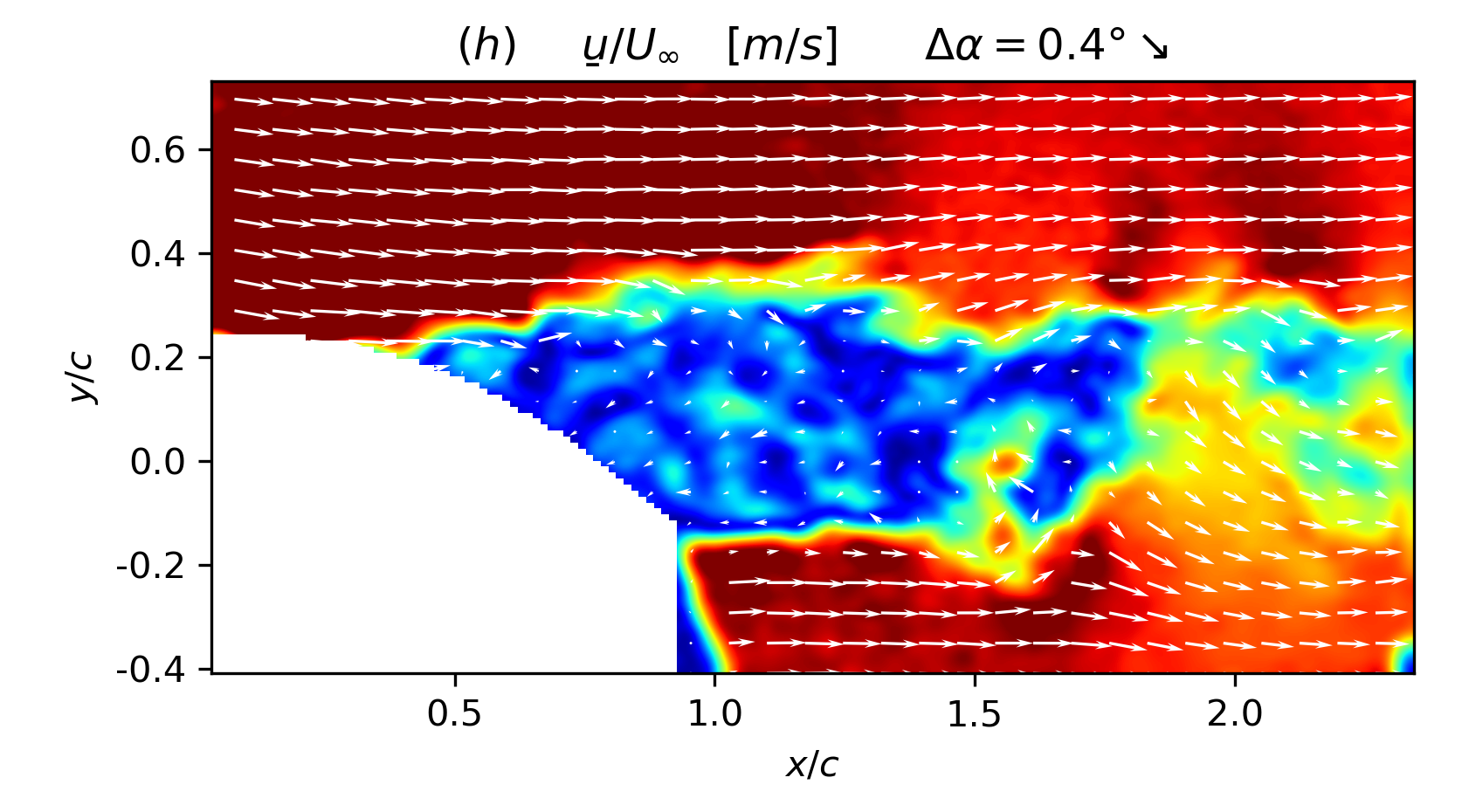}\tabularnewline
\multicolumn{2}{c}{\includegraphics[width=0.5\textwidth]{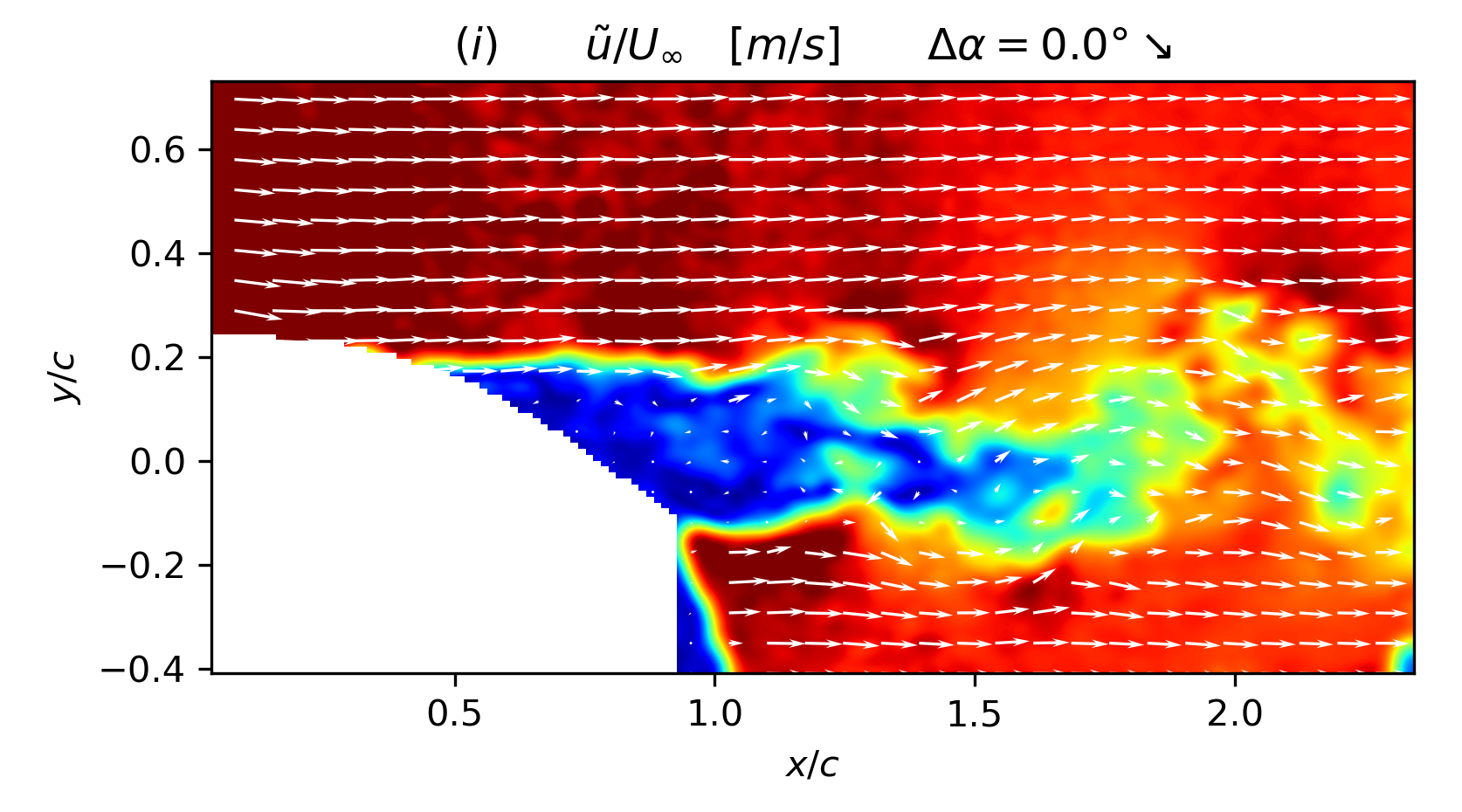}}\tabularnewline
\end{tabular} ~
\par\end{centering}
\begin{centering}
~
\par\end{centering}
\begin{centering}
~
\par\end{centering}
\caption{Instantaneous velocity fields superposed with isocontours of the velocity
modulus (i.e. $\sqrt{{\parallel u\parallel^{2}+\parallel v\parallel^{2}}}$
) at different $\Delta\alpha$ corresponding to points of the blade
oscillation given in figure \ref{fig:alpha_evolv} during the downstroke
phase (noted $\searrow$): (e) corresponds to a point at the maximum
amplitude of the blade oscillation cycle, (f) is an intermediate point,
(g) is a point just prior to the flow reattachment, (h) is a point
just after the flow reattachment and (i) is a point similar at the
lowest $\Delta\alpha$ of the downstroke phase \label{fig:Instantaneous-velocity-fields-1}}
\end{figure*}

One period of the blade oscillation relative
angle, $\Delta\alpha$, is extracted using the blade contour mask
from PIV images as explained in section \ref{subsec:strip detection}
(see figure \ref{fig:alpha_evolv}). The time duration $T$ and the
amplitude of the blade oscillation were chosen to include the flow
separation phenomena for quasi-static stall conditions, as previously
described in section \ref{subsec:blace_profile}. Points of interest
within this oscillating period are marked with letters from (a) to
(i) and the corresponding instantaneous vector fields are presented
in figures \ref{fig:Instantaneous-velocity-fields} and \ref{fig:Instantaneous-velocity-fields-1}.
At the beginning of the oscillating period, $\varDelta\alpha=0\text{\textdegree\ and t/T=0},$
the flow is slightly separated at the trailing edge of the profile
as can be seen in figure \ref{fig:Instantaneous-velocity-fields}a.
From point (a) to (c), corresponding to a positive blade incidence
variation, the separation point moves gradually from the trailing
edge to the leading edge of the profile and the wake width increases
accordingly as illustrated from \ref{fig:Instantaneous-velocity-fields}a
to \ref{fig:Instantaneous-velocity-fields}b. From point (c) to point
(d) the separation point suddenly moves towards the leading edge with
a corresponding increase of the wake width, until the flow is fully
separated over the aerodynamic profile (see figure \ref{fig:Instantaneous-velocity-fields}c
and d). This last phenomena is ten times faster than the previous
one and is clearly related to the stall phenomena. From point (d)
to point (e), the flow past the blade can clearly be considered as
a wake flow with a separation that occurs on both sides of the blade,
the leading and trailing edges (see figure \ref{fig:Instantaneous-velocity-fields}d
and e). 

From point (e) to (g), despite the progressive decrease of the adverse
pressure gradient on the suction side of the blade through a negative
variation of the blade incidence during 0.3 seconds, the flow remains
fully separated (see figure \ref{fig:Instantaneous-velocity-fields-1}
e, f and g). From point (g) to point (h), corresponding to a duration
of $\Delta t=0.02s$, the separation point suddenly moves back towards
the trailing edge. Again, this phenomena is ten times faster than
the time duration from (e) to (g) for which the blade incidence is
progressively decreasing (see figure \ref{fig:Instantaneous-velocity-fields-1}
g and h). From point (h) to (i), the separation point is back to its
initial state (see figure \ref{fig:Instantaneous-velocity-fields-1}
h and i). The stall and reattachment instants are defined respectively
as $t_{stall}^{ref}(ic)=(t_{c}+t_{d})/2$ and $t_{attach}^{ref}(ic)=(t_{g}+t_{h})/2$
with $t_{c}$, $t_{d}$, $t_{g}$ and $t_{h}$ the instants (c), (d),
(g) and (h) extracted from $ic=1$ to $N_{cycle}$, $N_{cycle}=18$
being the total number of instantaneous oscillation cycles. They will
be used in the following as a reference for the flow separation/reattachement
detection methods of section \ref{subsec:detection-methods}. 

It should be emphasize that the separation/detachment phenomena has
a time scale corresponding to $\sim10c/U_{\infty}$ in good agreement
with the theoretical work of Jones \citep{jones_unsteady_1940}, with
a separation/detachment location which occurs within one third of
the blade chord from the leading edge. 

\begin{figure}
\begin{centering}
\includegraphics[width=0.5\textwidth]{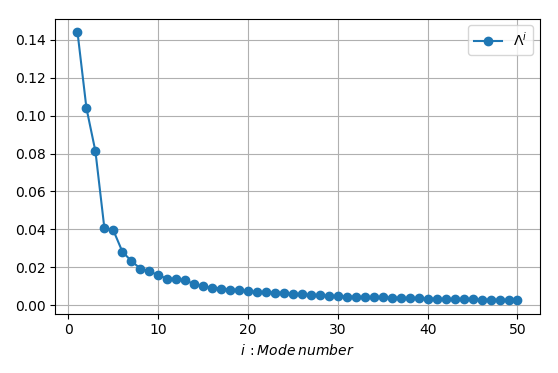} 
\par\end{centering}
\caption{Energy content of each of the first 50 POD modes \label{fig:pod_convergence}}
\end{figure}

To characterize further the coherent structure organization during
this blade oscillation cycle, a POD analysis is performed from a database
coming from a higher PIV acquisition rate, 1600Hz. All vector fields
of the blade oscillation cycles are used for the computation of the
temporal autocorrelation coefficient $C$ (see section \ref{subsec:POD}),
corresponding to 2000 snapshots. The convergence of the resulting
POD decomposition, in term of the relative energy content with modes,
is presented in figure \ref{fig:pod_convergence} using the following
definition :

\[
\Lambda_{i}=\frac{\lambda_{i}}{\sum_{j=1}^{^{N}}\lambda_{j}}
\]

where $N$ is the number of modes and $\lambda_{i}$ the eigenvalue
of the $i$th-mode.

\begin{figure*}
\begin{centering}
\begin{tabular}{cc}
\multicolumn{2}{c}{\includegraphics[width=0.5\textwidth]{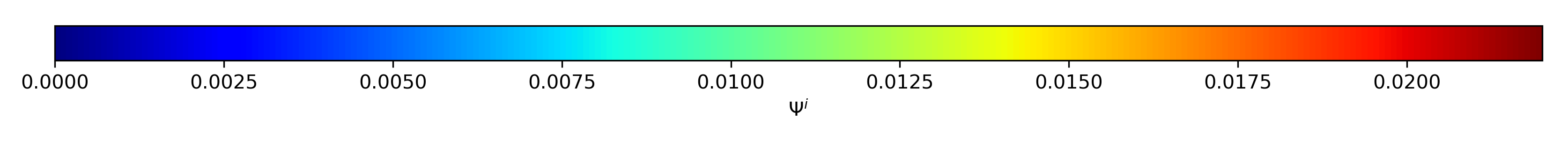}}\tabularnewline
a)\includegraphics[width=0.5\textwidth]{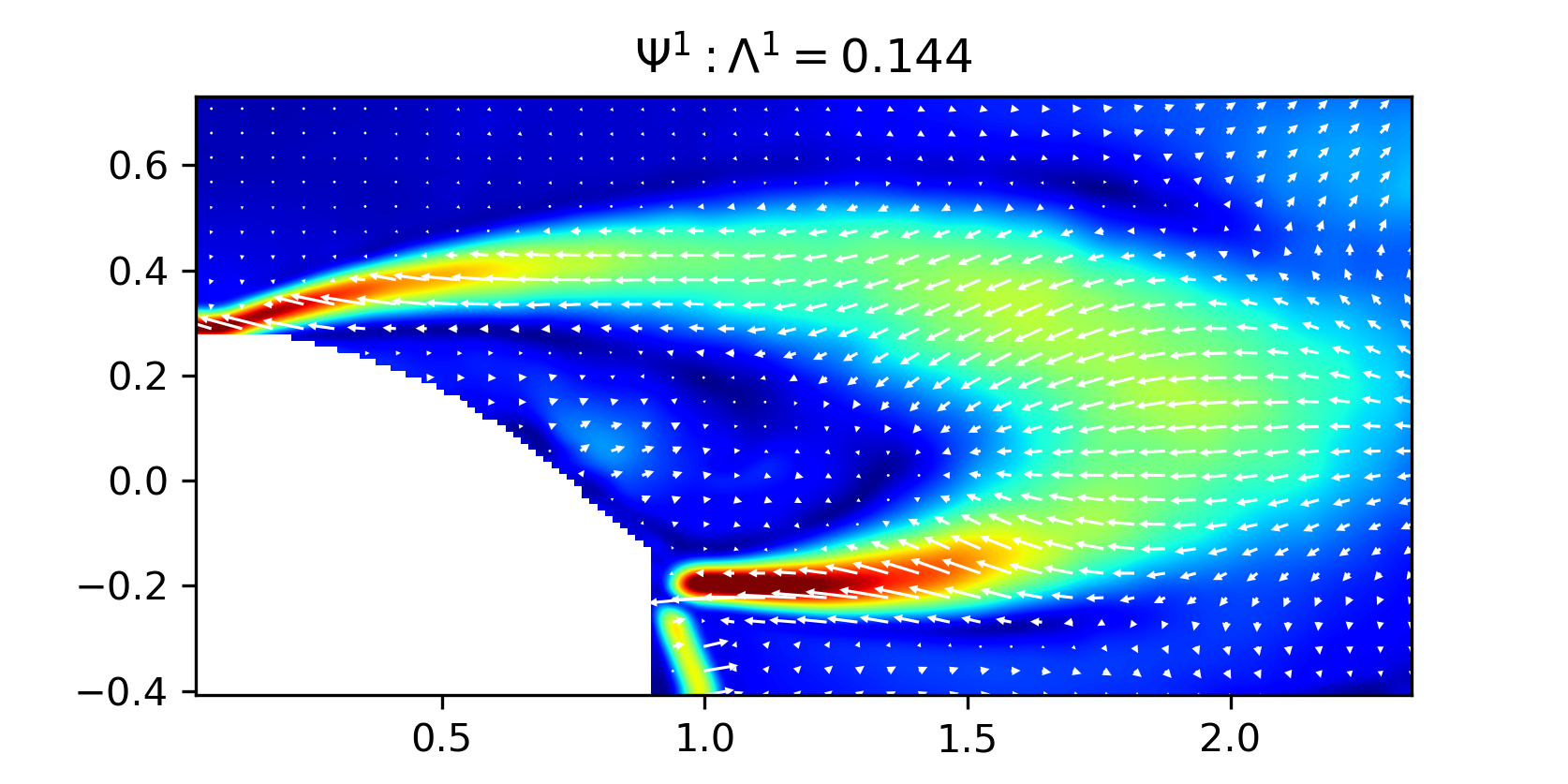} & b)~\includegraphics[width=0.5\textwidth]{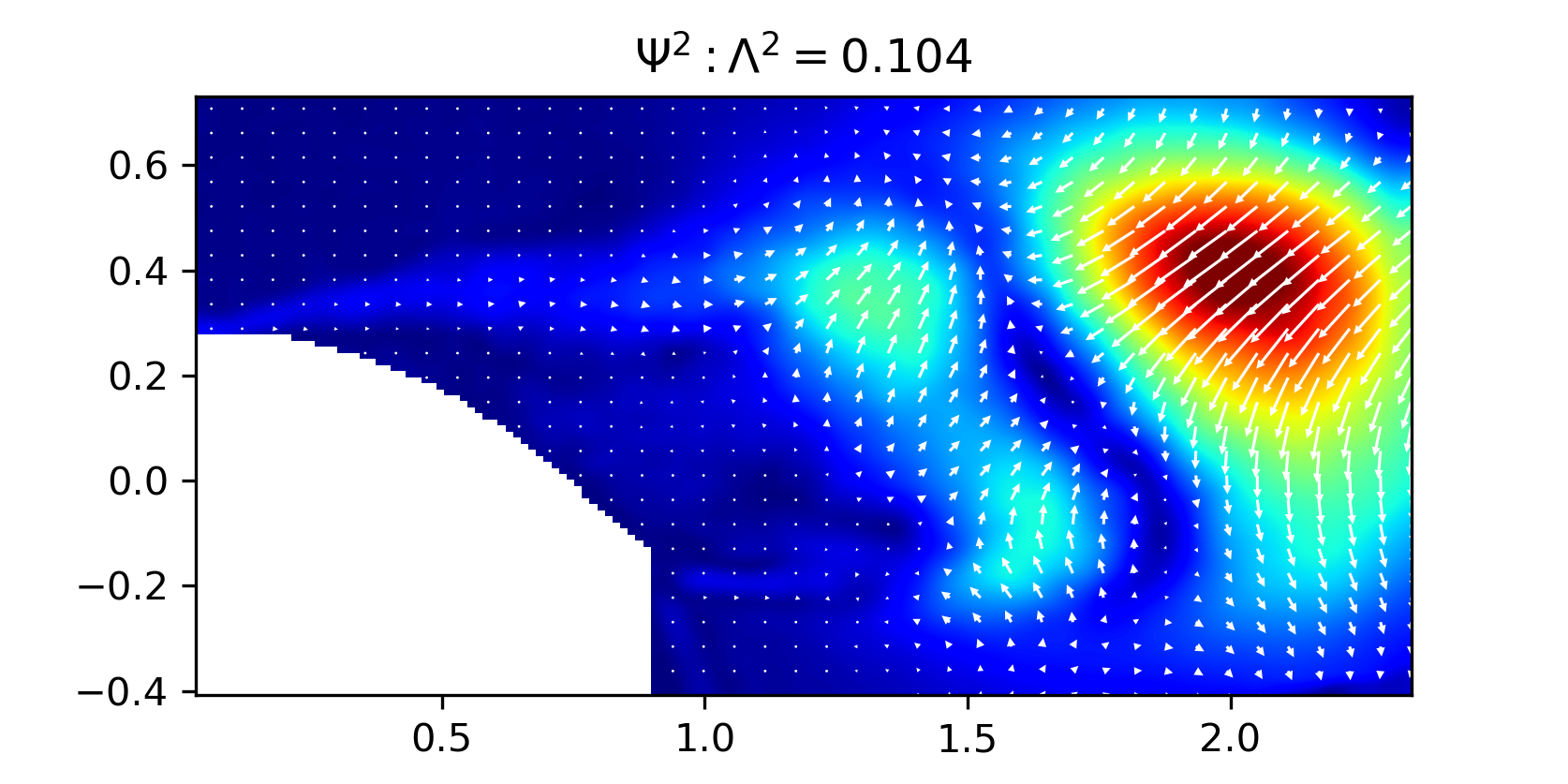}\tabularnewline
\hline 
c)\includegraphics[width=0.5\textwidth]{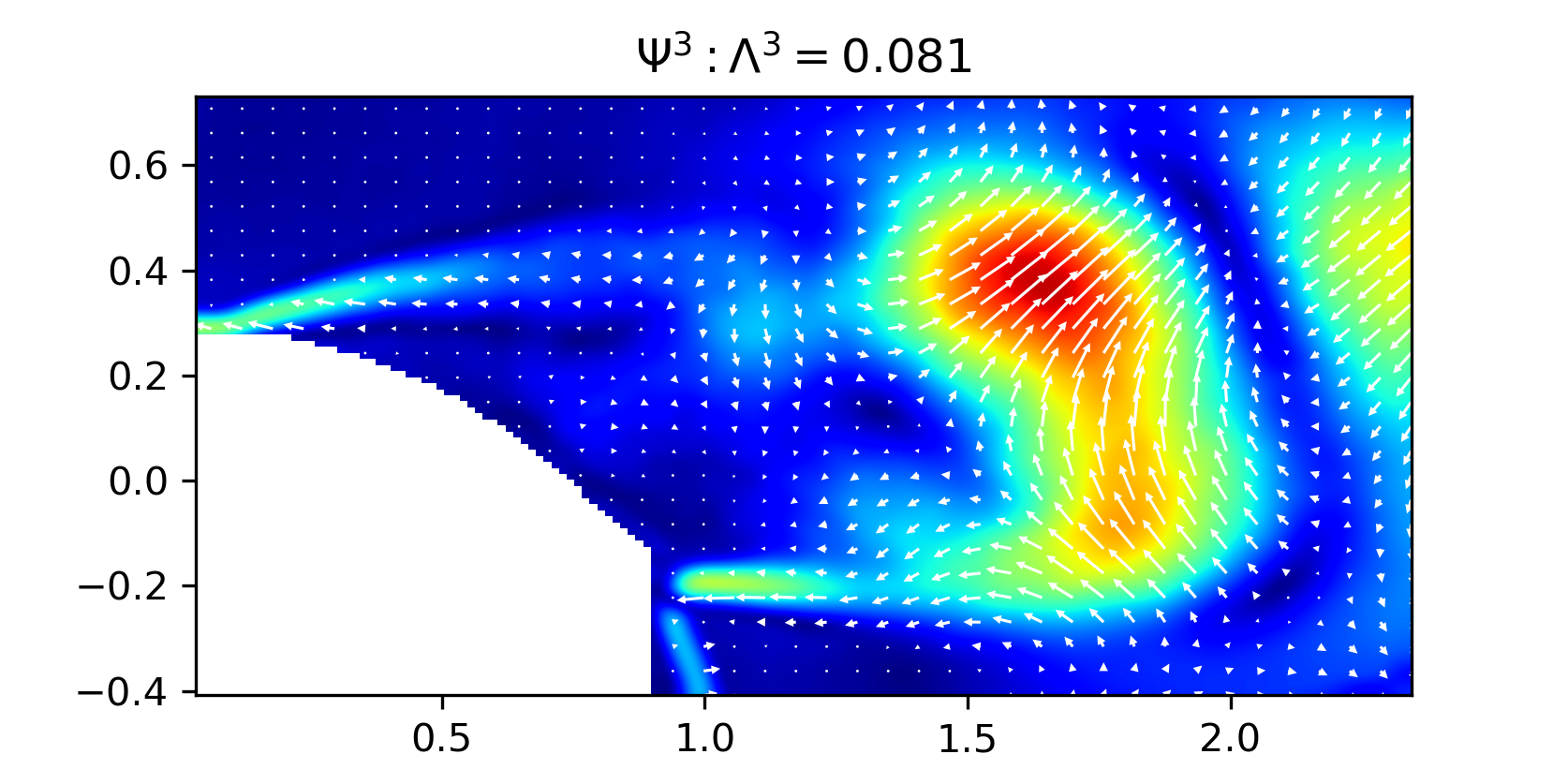} & ~d)\includegraphics[width=0.5\textwidth]{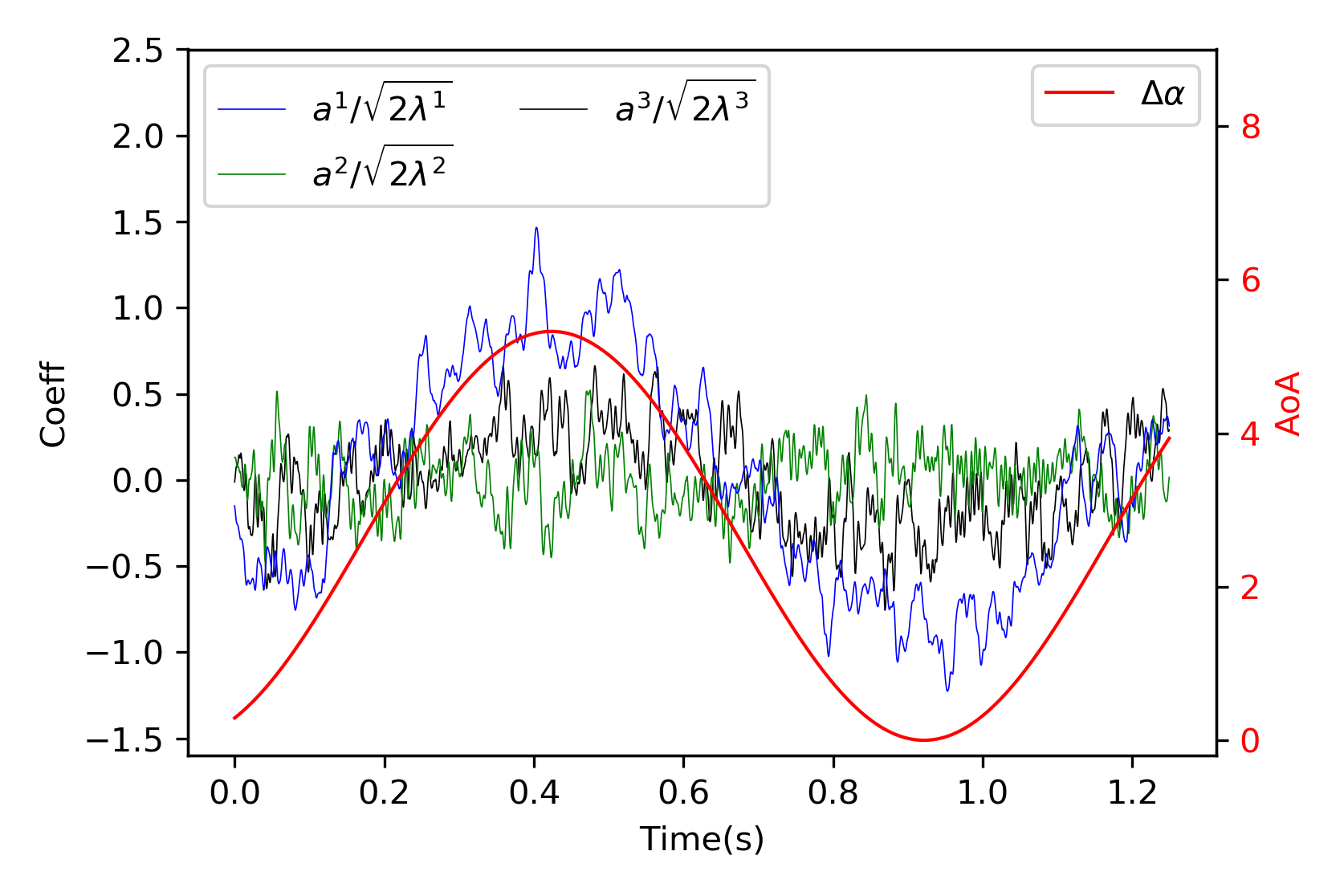}\tabularnewline
\hline 
\end{tabular} 
\par\end{centering}
\caption{POD decomposition: a), b) and c) represents the eigenvectors vector
field, $\Psi_{i}^{n},i=1,2$ , of the first three modes respectively
($n=1,2,3$) with isocontours of its modulus superimposed, the associated
energy content of the $n$ th mode (i.e. $\Lambda^{n})$ being written
in the title, d) represents the corresponding temporal coefficients
scaled with their energy content \label{fig:POD_modes_sspenon}}
\end{figure*}

\begin{figure}
\begin{centering}
\includegraphics[width=0.7\textwidth]{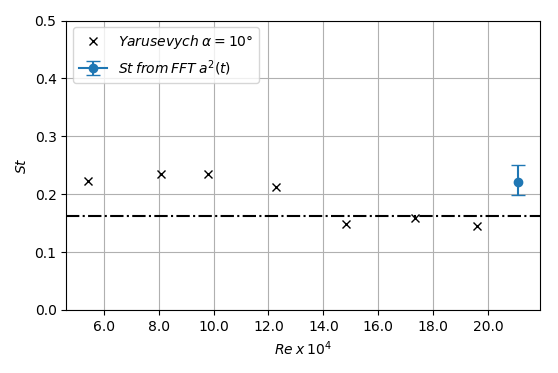} 
\par\end{centering}
\caption{Strouhal number values extracted from \citep{yarusevych_vortex_2009}
and from the FFT of the temporal mode, $a^{2}(t)$, of the POD decomposition
\label{fig:St}}
\end{figure}

As highlighted from figure \ref{fig:pod_convergence}, the dominant
modes in term of energy content are the three first POD modes, with
around 14\% of kinetic turbulent energy for the first mode, 10\% for
mode 2 and of 8\% for mode 3. These three modes are represented in
figure \ref{fig:POD_modes_sspenon} using the spatial modes, $\underline{\Psi}^{n}(\underline{x})$
with $n=1,2,3$, together with the temporal modes scaled with the
associated energy content, $a^{n}(t)/(2\lambda^{n})$ with $n=1,2,3$.
The first mode is phased with the blade oscillation period and clearly
captures variations of the mean velocity deficit in the wake due to
these oscillations. The second and third modes exhibit structures
in the wake which could be associated to the vortex shedding organization,
typically found in the wakes of bluff bodies. Following the work of
Yarusevych et al (2009) \citep{yarusevych_vortex_2009}, the Strouhal
number $St=f_{s}d/U_{\infty}\sim0.22$ is extracted, with $f_{s}$
the peak frequencies from the FFT of temporal modes, $a^{n}(t)/(2\lambda^{n})$
with $n=2,3$, and $d$ a measure of the wake width using the vertical
distance between the two local maximum of the $r.m.s$ of the streamwise
velocity at $x/c=1.25$ . This Strouhal number is of the same order
of magnitude of the one found by Yarusevych et al (2009) \citep{yarusevych_vortex_2009}
behind the wake of a NACA 0025 airfoil at the angle of attack of 10\textdegree{}
and clearly assess the link of these modes to the vortex shedding
organization behind the blade wake (see figure \ref{fig:St}). 

\subsection{Detection methods }

\label{subsec:detection-methods}

To be able to study the ability of the strip to detect the instants
of the flow separation/reattachment, three robust detection methods
were applied to the flow field obtained from the TR-PIV measurements:
\begin{itemize}
\item Method 1: using of the tangential instantaneous velocity component
in the direction perpendicular to the surface as introduced by \citep{de_gregorio_flow_2007}
\item Method 2: using the extraction of vortices in the shear layers as
explained in section \ref{subsec:vortexID}
\item Method 3: using the POD decomposition introduced in section \ref{subsec:POD}
\end{itemize}
In the perspective of using these sensors for real time control/monitoring
purposes, the application of these methods to instantaneous signals
is preferred. For each method a stall/reattachment criteria is defined,
corresponding to a zero crossings criteria explained in the introduction
of the first method (next section). 

\subsubsection{Method 1 \label{subsec:method1}}

\begin{figure}
\begin{centering}
\begin{tabular}{c}
a) \includegraphics[width=0.7\textwidth]{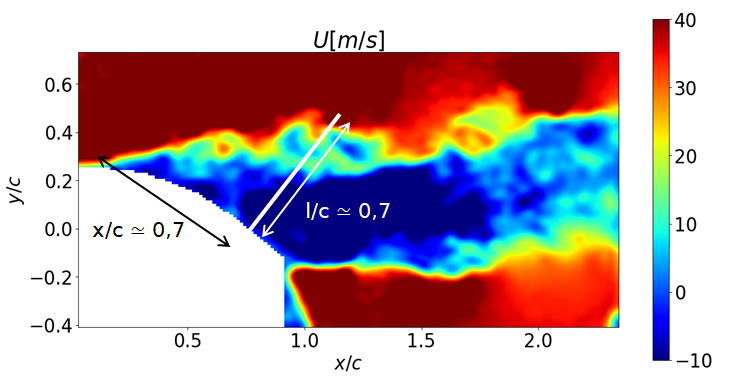}\tabularnewline
b) \includegraphics[width=0.7\textwidth]{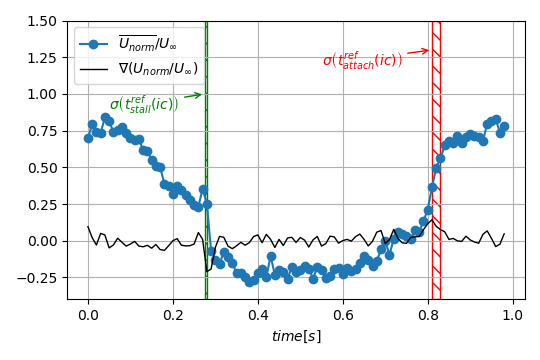}\tabularnewline
c)\includegraphics[width=0.7\textwidth]{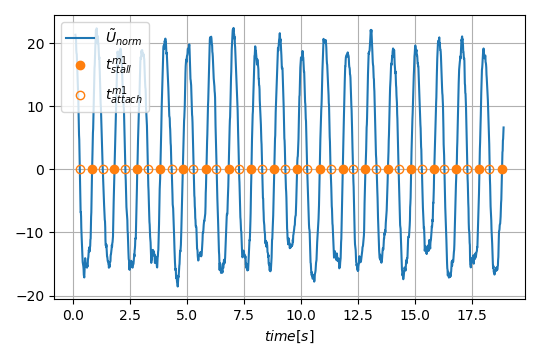} \tabularnewline
\end{tabular}
\par\end{centering}
\caption{First method to detect the flow stall/reattachment instants: a) location
and direction of integration line used to compute $U_{norm}(t_{i})$(i.e.
white bar on the blade) reported on isocontours of the velocity modulus
from PIV measurements, b) The phased-averaged signal $\overline{U_{norm}}$
with its gradient $\nabla\overline{U_{norm}}$ , c) results of the
zero-crossing method to extract the flow stall/reattachment instants
using the first method. The filled circle symbols correspond to stall
instants, $t_{stall}^{m1}(ic)$ , and void circle symbols corresponds
to reattachment instants, $t_{attach}^{m1}(ic)$. \label{fig:line_Unorm}}
\end{figure}

For the first method, the flow separation/reattachment instants are
detected for each oscillating cycle using the normal profile of the
instantaneous streamwise velocity component at a position corresponding
to the attached strip location $x_{strip}=x/c\simeq0.7$, $U_{norm}(t_{i},x_{strip,}y_{b})$
with $i$ the number of snapshots and $y_{b}$ the direction normal
to the blade surface. The chosen line location is presented in white
on the figure \ref{fig:line_Unorm}a. The normal profile is then reduced
to a single value by averaging in the normal direction, $U_{norm}(t,x_{strip})=\int_{y_{b}=0}^{l/c}U_{norm}(t,x_{strip},y_{b})dy_{b}$
with $l/c\simeq0.7$ the normalized integration length in the normal
direction, chosen so that each instant (or each angle of incidence)
corresponds to one value of this normal velocity. Different positions
and integration lengths have been explored with no significant influence
on the results. This could be partly explained from the fact that
PIV measurements do not capture the boundary layer gradient from the
aerodynamic surface. The phased average of the obtained $U_{norm}(t)$
signal, $\overline{U_{norm}}$, is presented in figure \ref{fig:line_Unorm}b
together with its gradient for further understanding of this detection
method.

For low angles of incidence $\overline{U_{norm}}/U_{\infty}\simeq1$,
meaning $\overline{U_{norm}}$ is close to the free-stream velocity
which corresponds to an attached flow state over the aerodynamic surface.
Similarly, for the large angles of incidence $\overline{U_{norm}}/U_{\infty}$
is negative, bringing to light the reverse flow above the profile
and thus the flow separation state. The time window width marked by
green and red hatched areas corresponds to the standard deviation
$\sigma(t_{(stall-or-attach)}^{ref}(ic)-ic.T)$ centered on the averaged
of the reference instants extracted from the instantaneous velocity
fields of section\ref{subsec:baseline_flow}, $\overline{t_{stall}^{ref}(ic)}$
and $\overline{t_{attach}^{ref}(ic)}$. Gradient peaks are close to
these reference instants which constitute a first validation of the
method. To extract the stall or reattachment instants from the instantaneous
$U_{norm}(t,x_{trip})$ signal, defined respectively as $t_{stall}^{m1}(ic)$
and $t_{attach}^{m1}(ic)$, it is first smoothed using a centered
moving average algorithm using a filter width of 21 time steps.Then,
a zero-crossing criteria is applied. This criteria uses the $U_{norm}(t)$
signal removed by its mean value, $\widetilde{U}_{norm}(t)$, so that
sudden variations of the signal are located where the fluctuating
signal is crossing the x-axis. Finally, the sign of the gradient,
$sign(\nabla U_{norm})$, is used to discriminate stall instants from
reattachment instants, $t_{stall}^{m1}(ic)$ and $t_{attach}^{m1}(ic)$
(see figure \ref{fig:line_Unorm}c). This zero-crossing method will
be also used for the detection methods 2 and 3 that follows.

\begin{figure}
\begin{centering}
\includegraphics[width=0.7\textwidth]{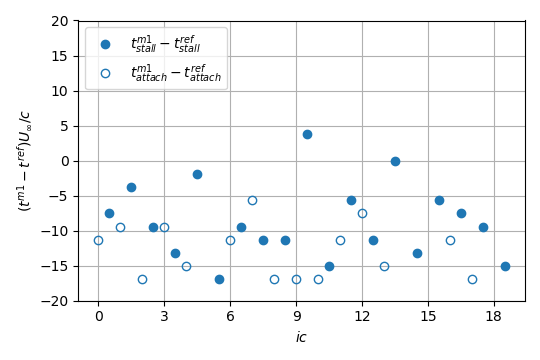} 
\par\end{centering}
\caption{Normalized delay of the stall and reattachment detected instants using
$U_{norm}(t)$ (method 1) \label{fig:Unorm_ecarts}}
\end{figure}

The resulting detected instants, $t_{stall}^{m1}(ic)$ and $t_{attach}^{m1}(ic)$
are compared to the reference instants extracted from the instantaneous
velocity fields of section\ref{subsec:baseline_flow}, $t_{stall}^{ref}(ic)$
and $t_{attach}^{ref}(ic)$ (see figure \ref{fig:Unorm_ecarts}).
The first observation is that the stall and reattachment instants
are detected earlier in average, $\frac{-8.6c}{U_{\infty}}$ and $\frac{-15c}{U_{\infty}}$
chord time respectively (or 2.5 to 4 time steps), when using this
first detection method. Then, a certain dispersion exist in the detected
instants that can be quantified using the standard deviations, $\frac{\sigma(t_{stall}^{m1}(ic)-ic.T)}{c/U_{\infty}}=3.3$
and $\frac{\sigma(t_{attach}^{m1}(ic)-ic.T)}{c/U_{\infty}}=5.0$.
It is found of the same order of magnitude as the reference case,$\frac{\sigma(t_{stall}^{ref}(ic)-ic.T)}{c/U_{\infty}}=2.1$
and $\frac{\sigma(t_{attach}^{ref}(ic)-ic.T)}{c/U_{\infty}}=7.0$.
Also, knowing the time resolution is $3.5U_{\infty}/c$ , this dispersion
should be attributed to the time accuracy of the dataset. 

\subsubsection{Method 2}

\begin{figure}
\begin{centering}
\begin{tabular}{c}
a) \includegraphics[width=0.8\textwidth]{Figure_4}\tabularnewline
b)\includegraphics[width=0.7\textwidth]{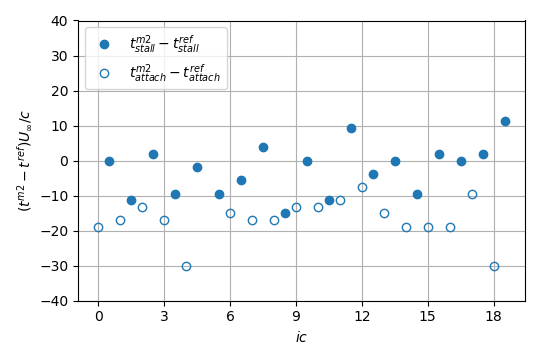}\tabularnewline
\end{tabular}
\par\end{centering}
\caption{Second method to detect the flow stall/reattachment instants: a) The
phased-averaged signal $\overline{W(t)}$ with its gradient $\nabla\overline{W(t)}$,
b) results of the zero-crossing method to extract the flow stall/reattachment
instants using the second method, $t_{stall}^{m1}(ic)$ and $t_{attach}^{m1}(ic)$,
compared to reference instants. The time window width marked by green
and red hatched areas corresponds to the standard deviation $\sigma(t_{(stall-or-attach)}^{ref}(ic)-ic.T)$
centered on the averaged of the reference instants extracted from
the instantaneous velocity fields of section\ref{subsec:baseline_flow},
$\overline{t_{stall}^{ref}(ic)}$ or $\overline{t_{attach}^{ref}(ic)}$.
The filled circle symbols correspond to stall instants, and void circle
symbols corresponds to reattachment instants. \label{fig:W_pa}}
\end{figure}

Another flow separation detection method is introduced with this time
a criteria associated with physical grounds. Indeed, it is using the
vertical distance between identified vortices in the separated shear
layers forming the blade wake width, directly related to the flow
separation location on the aerodynamic surface \citep{yarusevych_vortex_2009}
(see \ref{subsec:vortexID} on the vortex identification method).
The wake width is defined as : 

\begin{equation}
W(t)=\mid\frac{1}{N_{clock}(t)}\sum_{n=1}^{N_{clock}(t)}y_{n}(t)-\frac{1}{N_{anti-clock}(t)}\sum_{m=1}^{N_{anti-clock}(t)}y_{m}(t)\mid
\end{equation}
 with subscripts $clock$ and $anti-clock$ corresponding to quantities
from the clockwise and anti-clockwise rotating vortices respectively
and $N$ the number of vortices identified at the time $t$. The obtained
signal can be phased averaged, $\overline{W(t)}$, as presented in
figure \ref{fig:W_pa}a. Gradient peaks of the $\overline{W(t)}$
signal are close to the reference instants, which standard deviation
is represented by green and red hatched areas. This constitute a first
validation of the method. As for the first method, the zero-crossing
criteria is applied to the resulting filtered temporal evolution of
$W(t)$ to obtain stall and separated instants, $t_{stall}^{m2}(ic)$
and $t_{attach}^{m2}(ic)$. First results show that the mean detected
stall instant is closer to the reference than the first detection
method, i.e. $\frac{-2.5c}{U_{\infty}}$ (less than one time step),
while the reattachment instant is detected significantly earlier $\frac{-18c}{U_{\infty}}$.
The dispersion in the detected instant computed using the standard
deviation is of the order of 2 time steps for both the stall and the
reattachment , i.e. $\frac{\sigma(t_{stall}^{m2}(ic)-ic.T)}{c/U_{\infty}}=6.8$
and $\frac{\sigma(t_{attach}^{m2}(ic)-ic.T)}{c/U_{\infty}}=7.8$.
The increased dispersion in the detected stall instants compared to
the reference case should be attributed to the higher difficulty to
detect shear layer vortices for instants before stall, for which vortices
are smaller (i.e. within the spatial resolution of PIV measurements). 

\subsubsection{Method 3}

\begin{figure}
\begin{centering}
\includegraphics[width=0.7\textwidth]{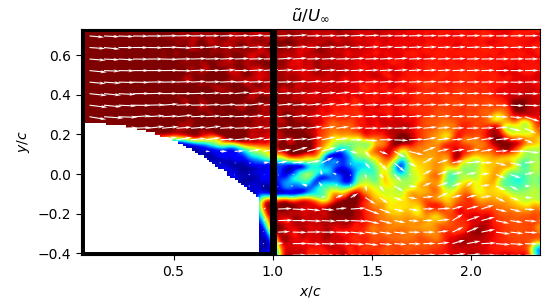} 
\par\end{centering}
\caption{Reduced field of view (black rectangle) used for the third detection
method using POD. \label{fig:FOV}}
\end{figure}

\begin{figure*}
\begin{centering}
\begin{tabular}{cc}
\multicolumn{2}{c}{\includegraphics[width=0.5\textwidth]{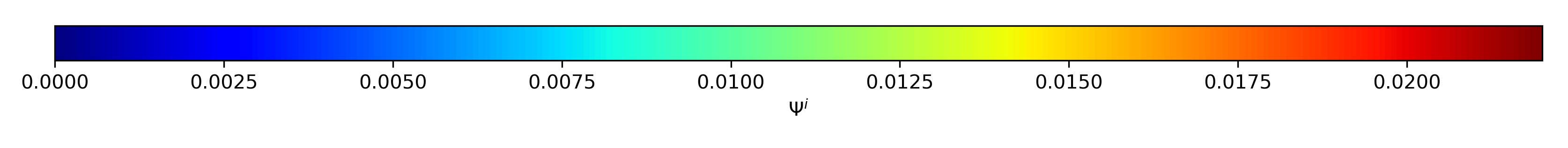}}\tabularnewline
a)\includegraphics[width=0.5\textwidth]{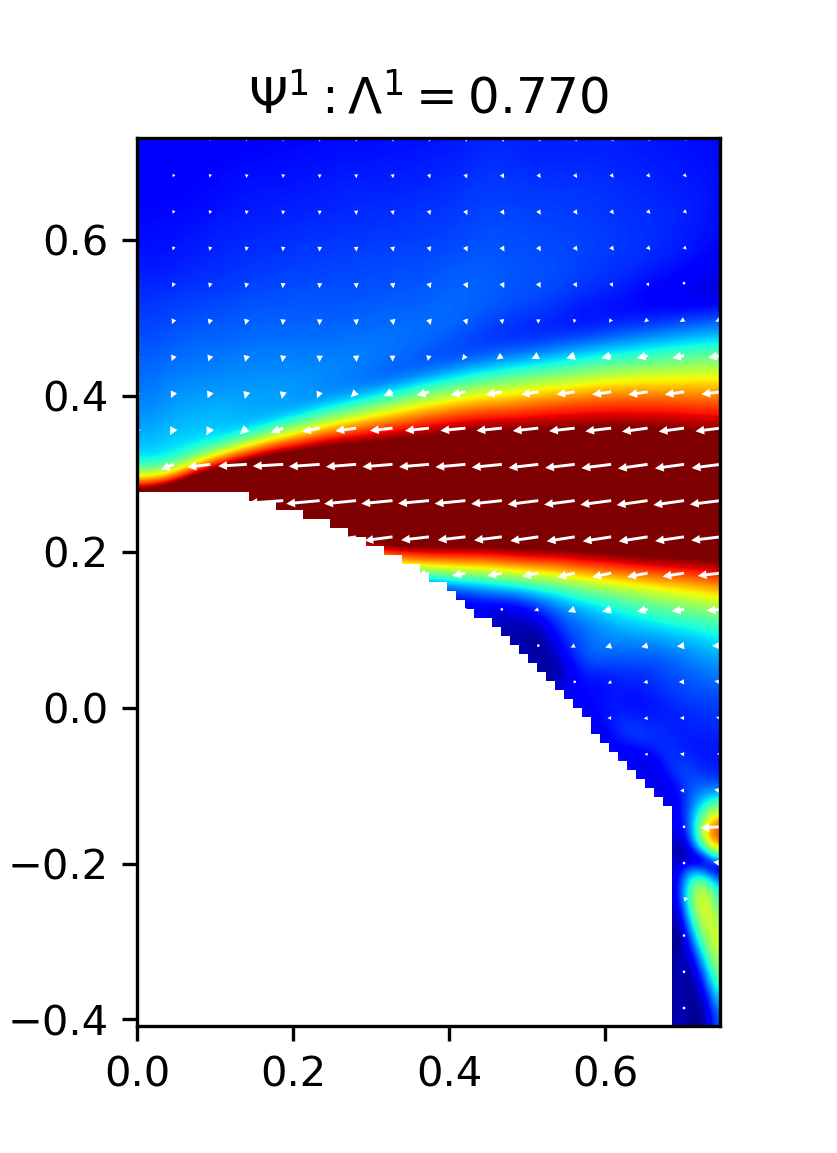} & b)\includegraphics[width=0.5\textwidth]{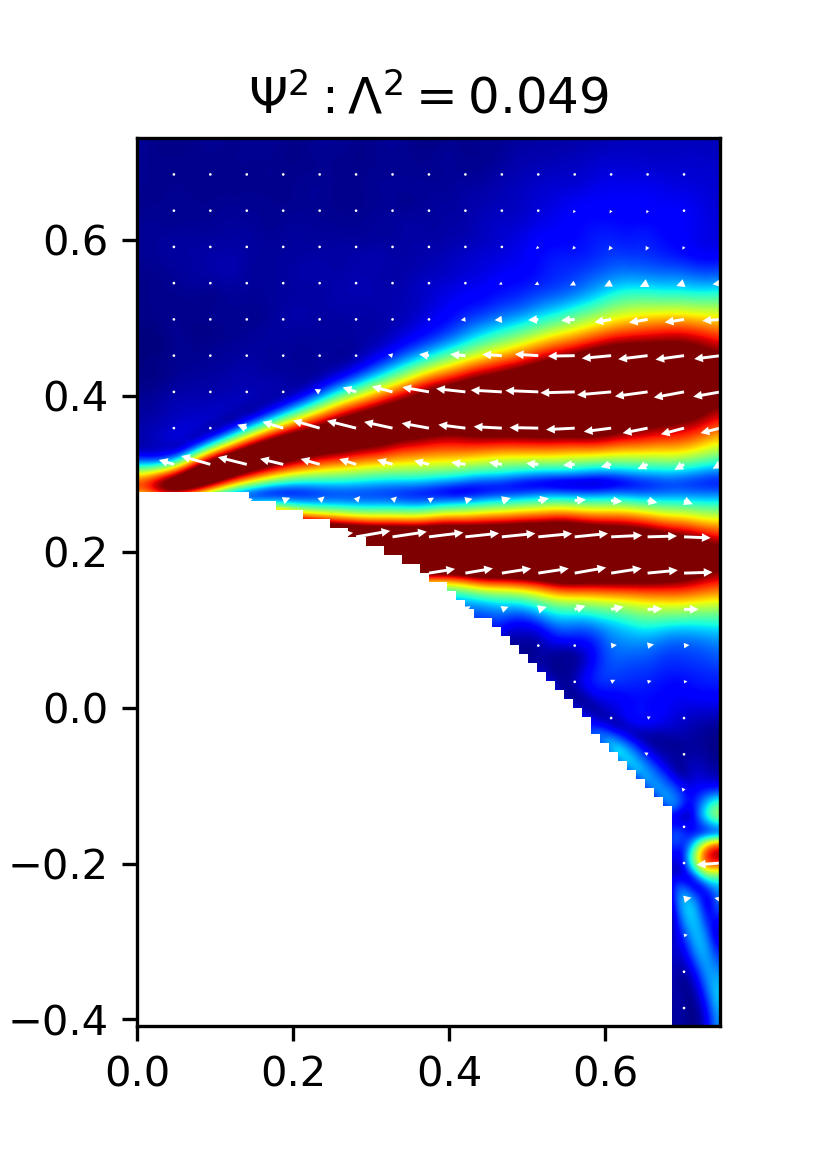}\tabularnewline
\multicolumn{2}{c}{c)\includegraphics[width=0.7\textwidth]{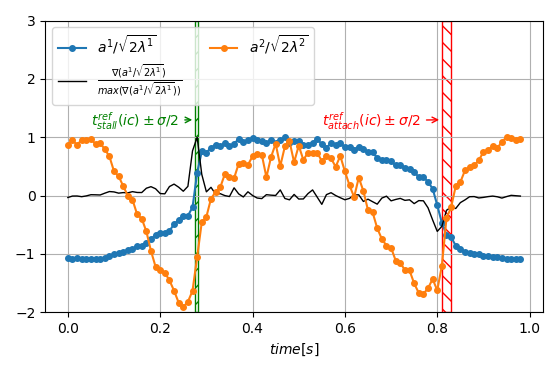}}\tabularnewline
\end{tabular} 
\par\end{centering}
\caption{Third method to detect the flow separation/reattachment instants:
a) and b) are isocontours of the eigenvectors field $\Psi_{i}^{n}$,
with $i=1,2$, of the $n$-th mode, , with isocontours of its modulus
superimposed. $\Lambda^{n}$ is the eigenvalue of the $n$-th mode,
representing the part of the turbulent kinetic energy in the mode.
c) represents the phase averaged of the corresponding temporal coefficients
scaled with their turbulent kinetic energy content ($a^{n}(t)/\sqrt{2\lambda^{n}},n=1,2,$)
\label{fig:POD_modes_red_sx}. The time window width marked by green
and red hatched areas corresponds to the standard deviation $\sigma(t_{(stall-or-attach)}^{ref}(ic)-ic.T)$
centered on the averaged of the reference instants extracted from
the instantaneous velocity fields of section\ref{subsec:baseline_flow},
$\overline{t_{stall}^{ref}(ic)}$ or $\overline{t_{attach}^{ref}(ic)}.$}
\end{figure*}

These two previous methods provide an instantaneous detection of the
flow separation phenomena. To explore further the detection of these
instants, we choose to use another method based on statistics introduced
in section \ref{subsec:POD}. It was already used in the context of
wind energy for the analysis of the dynamic stall phenomena by Melius
et al (2016) \& Mulleners et al (2016) \citep{melius_dynamic_2016,mulleners_onset_2012}.
The chosen vector field for the present analysis focuses on the separated
shear layer dynamics rather than the wake dynamics from the initial
PIV field of view (see figure \ref{fig:FOV}). 2000 snapshots were
used with no distinction of the phase, which enables to extract the
flow separation state within the first POD modes as explained by \citep{melius_dynamic_2016,mulleners_onset_2012}.
As a first approach, the phased average of the two first POD modes
are presented in figure \ref{fig:POD_modes_red_sx} with temporal
coefficients $a^{1}(t)$ and $a^{2}(t)$ . The first mode of the eigenvector
field presented in figure \ref{fig:POD_modes_red_sx}a (i.e. $\Psi_{i}^{1},i=1,2$),
contains 77\% of the total turbulent kinetic energy (i.e. $\Lambda^{1}\sim0.77$)
and captures accelerations and deceleration of the flow over the profile
depending on the sign of the associated temporal coefficient $a^{1}(t)$.
The transitions between the accelerations (i.e. $a^{1}(t)$ < 0) and
deceleration (i.e. $a^{1}(t)$ > 0) phases is marked by abrupt variations
of amplitudes, which should be associated to instants of the stall
and the flow reattachment phenomenon. The second mode of the eigenvector
field presented in figure \ref{fig:POD_modes_red_sx}b (i.e. $\Psi_{i}^{2},i=1,2$),
contains much less turbulent kinetic energy (i.e. $\Lambda^{2}\sim0.049$
) and exhibits a shear layer with a shear direction that is changing
accordingly with the sign of its associated temporal coefficient $a^{2}(t)$.
This variation of shear may be associated to the passage of the famous
dynamic stall vortex created during unsteady variations of the angle
of incidence as pointed out by \citep{melius_dynamic_2016,mulleners_onset_2012}.
Interestingly, minimums of $a^{2}(t)$ occurs significantly ahead
of the flow separation/reattachment instants contrary to the first
mode. However, studying the ability of the e-telltale sensor to detect
dynamic stall vortex needs further investigations that won't be performed
in this work. The following will therefore focus on the first POD
mode.

The coefficient of the first mode $a^{1}(t)$ was also studied instantaneously
to compare with the other detection methods. A zero-crossing criteria
was applied to this instantaneous signal, leading to detected stall
and reattachment instants $t_{stall}^{m3}(ic)$ and $t_{attach}^{m3}(ic)$.
First results show these instants follows the trend of the first detection
method regarding the mean quantities, i.e. the detection occurs earlier
than the reference: $\frac{-6.2c}{U_{\infty}}$ and $\frac{-12c}{U_{\infty}}$
, and the dispersion is similar to the reference, i.e. $\frac{\sigma(t_{stall}^{m3}(ic)-ic.T)}{c/U_{\infty}}=1.7$
and $\frac{\sigma(t_{attach}^{m3}(ic)-ic.T)}{c/U_{\infty}}=5.7$. 

\begin{figure}
\begin{centering}
\includegraphics[width=0.7\textwidth]{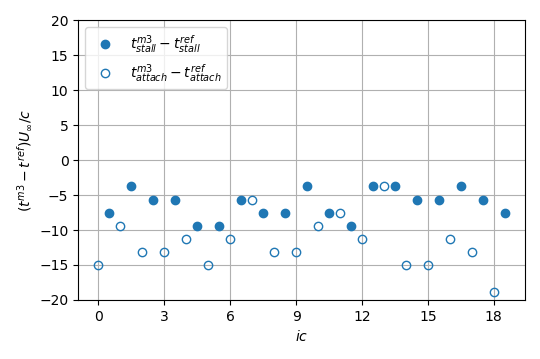}
\par\end{centering}
\caption{Results of the zero-crossing method to extract the flow separation/reattachment
instants using the third method. The filled circle symbols correspond
to stall instants, $t_{stall}^{m3}(ic)$, and void circle symbols
corresponds to reattachment instants, $t_{attach}^{m3}(ic)$\label{fig:a1_ecarts}}
\end{figure}

\subsection{Ability of the sensor to detect flow separation }

\begin{figure}
\centering{}\includegraphics[width=0.8\textwidth]{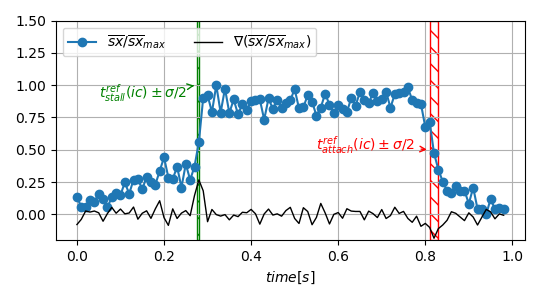}\caption{The evolution of dimensionless phase averaged streamwise coordinate
of the center of the strip, $sx/sx_{max}$, during the oscillation
cycle (blue dotted line) together with its gradient (black line) .
The time window width marked by the red hatched area corresponds to
the standard deviation $\sigma(t_{stall}^{ref}(ic)-ic.T)$ centered
on the averaged $\overline{t_{stall}^{ref}(ic)}$ . The time window
width marked by the green hatched area corresponds to the standard
deviation value $\sigma(t_{attach}^{ref}(ic)-ic.T)$ centered on the
phase averaged $\overline{t_{attach}^{ref}(ic)}$. . \label{fig:sx}}
\end{figure}

Detection methods 1, 2 and 3 using TR-PIV measurements will be compared
to the detection method using the e-TellTale sensor. For that purpose,
the phase averaged strip position, $\overline{sx(t)}$, is detected
from image processing as explained in section \ref{subsec:strip detection}
and presented in figure \ref{fig:sx} together with the time averaged
standard deviation of stall and reattachment instants detected from
the instantaneous flow field (i.e. green and red hatched areas respectivelly).
It is observed that the position of the strip during the oscillation
cycle is characterized by two sudden changes, revealed with the gradient
peaks, in very good agreement with the stall and reattachment instants
observed with the instantaneous flow field. This is a first validation
of the e-TellTale sensor to detect stall and reattachment instants. 

\begin{figure}
\begin{centering}
\includegraphics[width=0.7\textwidth]{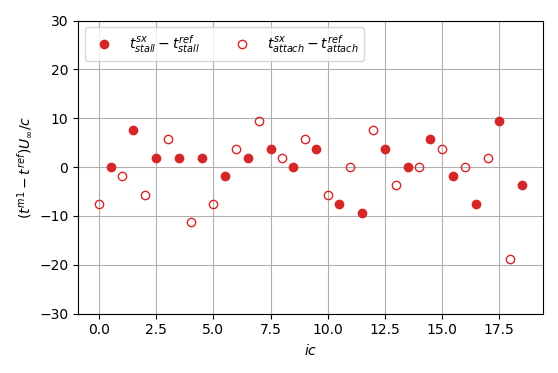}
\par\end{centering}

\caption{Comparison of the detected instants between the three methods using
the instantaneous velocity fields and the position of the strip.\label{fig:sx_ecarts}}
\end{figure}

To characterize further the detected instants from the movement of
the strip, the zero-crossing criteria is applied to the instantaneous
signal of the position of the strip, $sx(t)$. Resulting stall and
reattachment instants removed by the reference instants, $t_{stall}^{sx}(ic)-t_{stall}^{ref}(ic)$
and $t_{attach}^{sx}(ic)-t_{attach}^{ref}(ic)$, are plotted in figure
\ref{fig:sx_ecarts}. As highlighted here, the mean value is very
close to the reference (i.e. close to zero). Furthermore, when compared
to the detected instants from the three other methods summurized in tabular
\ref{tab:methods_compar_stats-1}, the e-TellTale detection method
presents the smallest delay to the reference, i.e. $\frac{0.5c}{U_{\infty}}$
and $\frac{-1.2c}{U_{\infty}}$. Futhermore, the dispersion difference
with the reference is within one time step, i.e. $3.5U_{\infty}/c$,
from the reference case, i.e. $\frac{\sigma(t_{stall}^{sx}(ic)-ic.T)}{c/U_{\infty}}=5.2$
and $\frac{\sigma(t_{attach}^{sx}(ic)-ic.T)}{c/U_{\infty}}=3.3$,
which should be attributed to the time accuracy.

\begin{table}
\begin{centering}
\begin{tabular}{>{\centering}p{0.1\textwidth}>{\centering}p{0.2\textwidth}>{\centering}p{0.2\textwidth}>{\centering}p{0.2\textwidth}>{\centering}p{0.2\textwidth}}
\cline{2-5} 
 & Mean of delays between stall detection with the different methods
and the reference (a) & Mean of delays between reattachment detection with the different methods
and the reference (b) & Standard deviation of the detected instants of stall (c) & Standard deviation of the detected instants of reattachment (d)\tabularnewline
\hline 
\hline 
Reference &  &  & 2.1 & 7.0\tabularnewline
\hline 
Method1 & -8.6 & -15 & 3.3 & 5.0\tabularnewline
\hline 
Method2 & -2.5 & -18 & 6.8 & 7.8\tabularnewline
\hline 
Method3 & -6.2 & -12 & 1.7 & 5.7\tabularnewline
\hline 
Epenon & 0.5 & -1.2 & 5.2 & 3.3\tabularnewline
\hline 
\end{tabular}
\par\end{centering}
\centering{}a) : $\frac{\sum_{ic=1}^{N_{cycle}}(t_{stall}^{mj}(ic)-t_{stall}^{ref}(ic))}{N_{cycle}c/U_{\infty}}$\enskip{}
b):$\frac{\sum_{ic=1}^{N_{cycle}}(t_{attach}^{mj}(ic)-t_{attach}^{ref}(ic))}{N_{cycle}c/U_{\infty}}$\enskip{}
c)$:\frac{\sigma(t_{stall}(ic)-ic.T)}{c/U_{\infty}}$\enskip{} d):$\frac{\sigma(t_{attach}(ic)-ic.T)}{c/U_{\infty}}$\caption{Summarize of detected instants values using the three methods including:
averaged detected instants compared to averaged reference instants
and standard deviations . All times are expressed as chord times $(t_{c}=c/U_{\infty})$\label{tab:methods_compar_stats-1}}
\end{table}

\section{Conclusion}

The ability of an original e-TellTale sensor to detect flow separation/reattachment
instants during an oscillation of the angle of incident of a blade
section has been explored. For that purpose, a 2D NACA 65-421 blade
section equipped with a e-TellTale sensor at its trailing edge has
been set in the LHEEA aerodynamic wind tunnel. The blade was oscillating
around the stall angle to reproduce a constant shear inflow perturbations
in front of a rotating wind turbine blade at a chord Reynolds number
of $2.10^{5}$. Three methods to detect the flow separation/reattachment
instants have been successfully applied using Time-Resolved-PIV measurements
during the blade oscillation cycle. This includes two instantaneous
methods, the direct use of the tangential instantaneous velocity (method
1) and the instantaneous extraction of shear layer vortices (method
2), and one statistical method using POD (method 3). Method 1 and
3 were found equivalent, with an earlier detection of the stall/reattachment
instants (2 to 4 time step earlier) and a dispersion in the detection
similar as the reference case. Method 2, using an instantaneous vortex
detection method, presents a detection of the stall closer to the
reference that is however detected with a higher dispersion. This
should be attributed to the difficulty to extract vortices when the
flow is attached. The e-TellTale detection method presents the best
results with a detection of the reference instants within less than
one time step and a dispersion similar as the reference case. This
study demonstrates the ability of the e-TellTale sensor strip to detect
the instantaneous separation/reattachment dynamics over the blade.
What remains to be done is a link between this dynamic strip position
and the dynamic response of e-TellTale strain gauge signal. Also,
the methodology used in the present paper is intended to be used in
the extraction of other flow features over the blade surface such
as the well known dynamic stall vortex or the blade wake dynamics. 

\section*{Acknowledgement}

Authors would like to thanks Jean-Jacques Lasserre, Philippe Galtier
for the PIV Dantec equipment loan and their assistance during measurements.
We also would like to thanks Vincent Jaunet for his help during the
PIV acquisition. This work was partly carried out within the framework
of the WEAMEC, West Atlantic Marine Energy Community, and with funding
from the city of Nantes, the Pays de la Loire Region and Centrale
Nantes in France. 

\bibliographystyle{abbrv}
\bibliography{ArticlePIV}

\end{document}